\documentclass[10pt,journal,compsoc]{IEEEtran}
\ifCLASSOPTIONcompsoc
  \usepackage[nocompress]{cite}
\else
  \usepackage{cite}
\fi
\ifCLASSINFOpdf
\else
\fi
\usepackage{graphicx,fontawesome}
\usepackage{amsmath,amssymb,textcomp,subfigure,multirow,upgreek}

\usepackage{tabularx}
\usepackage{makecell}
\usepackage{booktabs}
\usepackage{colortbl} 
\usepackage{tabularray}
\usepackage{longtable}

\usepackage[linesnumbered,algo2e,boxed]{algorithm2e}

\usepackage{tikz}
\usepackage[edges]{forest}
\usepackage{gensymb,graphics,subfigure}
\usepackage{float}
\graphicspath{{Figure/}}
\usepackage{xcolor}
\definecolor{Purple}{HTML}{7666ec}
\definecolor{Green}{HTML}{248067}
\definecolor{Red}{HTML}{7c1823}
\definecolor{Yellow}{HTML}{f28e16}
\definecolor{Blue}{HTML}{1661ab}
\definecolor{Pink}{HTML}{c06f98}

\usepackage{hyperref}
\hypersetup{
  breaklinks=true,
  colorlinks=true,
  citecolor=cyan,
  urlcolor=red,
  linkcolor=black
}

\usepackage{enumitem}
\usepackage{ragged2e} 
\usepackage{cleveref}

\usepackage{epsfig}
\usepackage{graphicx}

\usepackage{algorithm}
\usepackage{algorithmic}
\usepackage{caption}

\usepackage[figuresright]{rotating}
\usepackage{color}

\usepackage{ragged2e}
\graphicspath{{Figures/}}

\usepackage{wrapfig}
\usepackage{enumitem}
\setitemize{noitemsep,topsep=0pt,parsep=0pt,partopsep=0pt}

\usepackage{utfsym}
\usepackage{arydshln}
\usepackage{amssymb}
\begin{document}
%
\title{Style Transfer: A Decade Survey}
\author{Tianshan Zhang \quad Hao Tang$^*$
	\IEEEcompsocitemizethanks{
	    \IEEEcompsocthanksitem  Tianshan Zhang is with the School of Dalian Jiaotong University, Dalian, P.R.China.
	    E-mail: plote5024@gmail.com \protect  
	 \IEEEcompsocthanksitem Hao Tang is with the School of Computer Science, Peking University, Beijing 100871, China. E-mail: haotang@pku.edu.cn \protect
        }
	\thanks{$^*$Corresponding author: Hao Tang.}
}

\IEEEtitleabstractindextext{%
\justify
\begin{abstract}
The revolutionary advancement of Artificial Intelligence Generated Content (AIGC) has fundamentally transformed the landscape of visual content creation and artistic expression. While remarkable progress has been made in image generation and style transfer, the underlying mechanisms and aesthetic implications of these technologies remain insufficiently understood. This paper presents a comprehensive survey of AIGC technologies in visual arts, tracing their evolution from early algorithmic frameworks to contemporary deep generative models. We identify three pivotal paradigms—Variational Autoencoders (VAE), Generative Adversarial Networks (GANs), and Diffusion Models—and examine their roles in bridging the gap between human creativity and machine synthesis. To support our analysis, we systematically review over 500 research papers published in the past decade, spanning both foundational developments and state-of-the-art innovations. Furthermore, we propose a multidimensional evaluation framework that incorporates Technical Innovation, Artistic Merit, Visual Quality, Computational Efficiency, and Creative Potential. Our findings reveal both the transformative capacities and current limitations of AIGC systems, emphasizing their profound impact on the future of creative practices. Through this extensive synthesis, we offer a unified perspective on the convergence of artificial intelligence and artistic expression, while outlining key challenges and promising directions for future research in this rapidly evolving field.
\end{abstract}

}

\maketitle

\IEEEdisplaynontitleabstractindextext

%
\IEEEpeerreviewmaketitle


%
%
%
%

\section{Introduction}

" I dream my painting, and then I paint my dream."
\begin{flushright}
– Vincent Van Gogh.
\end{flushright}

\textbf{Style transfer} seeks a mapping $\mathcal{F}:(I_c,I_s)\!\mapsto\!I_t$ that preserves the structural semantics of a content image $I_c$ while matching the stylistic statistics of a reference image $I_s$. Since Gatys\cite{gatys2015neural}, the field has leapt from slow optimisation to millisecond feed-forward generators and, most recently, diffusion and autoregressive (AR) pipelines capable of 4K resolution and fine-grained semantic control. Style transfer powers portrait retouching, real-time video stylisation, and 3-D asset creation for games and film.

Over a decade of research reveals three disruptive leaps: (i) Neural Style Transfer demonstrated that convolutional Gram statistics encode transferable style; (ii) real-time generators and adversarial training brought photoreal fidelity; (iii) diffusion and AR systems in 2022-2025 pushed both scale and controllability. Yet the grand challenge remains: how to trade perceptual fidelity for speed and diversity under limited compute.

\begin{figure*}[ht]
    \centering 
    \begin{tikzpicture}[
        scale=0.6, 
        every node/.style={scale=0.7},
        node text/.style={align=center}
    ]
        \draw[thick, ->, >=stealth] (0,0) -- (26,0);

        \foreach \x/\year in {1/2013,3/2014,5/2015,7/2016,9/2017,11/2018,13/2019,15/2020,17/2021,19/2022,21/2023,23/2024,25/2025,
        } {
            \draw[thick] (\x, -0.2) -- (\x, 0.2);
            \node[below] at (\x, -0.2) {\year};
        }
        

        \node[above, text width=3cm, align=center,text=Purple] at (1,1.2) {
            \textbf{VAE}\cite{kingma2013auto}
        };

        \node[above, text width=3cm, align=center,text=Green] at (3,1.2) {\textbf{GAN}\cite{goodfellow2014generative}};

        \node[below, text width=3cm, align=center,text=Yellow] at (5,-1.2) {\textbf{A Neural Algorithm\\of Artistic Style}\cite{gatys2015neural}};

        \node[above, text width=3cm, align=center,text=Yellow] at (7,1.2) {
            \textbf{Real-Time Style Transfer}\cite{johnson2016perceptual}
        };
        \node[above, text width=3cm, align=center,text=Blue] at (7,2.2) {
            \textbf{PixelRNN}\cite{oord2016pixelrecurrentneuralnetworks}
        };

        \node[below, text width=3cm, align=center,text=Purple] at (9,-1.2)
        {\textbf{Beta-VAE}\cite{higgins2017beta}};

        \node[below, text width=3cm, align=center,text=Green] at (9,-1.7)
        {\textbf{CycleGAN}\cite{zhu2017unpaired}};

        \node[below, text width=3cm, align=center,text=Green] at (11,2.2){\textbf{AttnGAN}\cite{xu2017attnganfinegrainedtextimage}};
        
        \node[below, text width=3cm, align=center,text=Green] at (11,1.7)
        {\textbf{BigGAN}\cite{child2019biggan}
        };


        \node[below, text width=3cm, align=center,text=Green] at (13,-1.2){\textbf{StyleGAN}\cite{karras2019style}};
        
        \node[below, text width=3cm, align=center,text=Purple] at (13,-1.7)
        {\textbf{VQ-VAE-2}\cite{razavi2019generating}
        };


        \node[below, text width=3cm, align=center,text=Red] at (15,2.2){\textbf{DDPM}\cite{ho2020denoising}};
        
        \node[below, text width=3cm, align=center,text=Red] at (15,1.7)
        {\textbf{DDIM}\cite{song2020denoising}
        };

        
        \node[below, text width=3cm, align=center,text=Purple] at (17,-1.2){\textbf{DALL-E}\cite{ramesh2021zeroshottexttoimagegeneration}};
        
        \node[below, text width=3cm, align=center,text=Green] at (17,-1.7)
        {\textbf{VQ-GAN}\cite{esser2021tamingtransformershighresolutionimage}
        };

        \node[below, text width=3cm, align=center,text=Green] at (17,-2.2)
        {\textbf{DAE-GAN}\cite{ruan2021daegandynamicaspectawaregan}
        };

        \node[below, text width=3cm, align=center,text=Yellow] at (17,-2.7){\textbf{CLIP}\cite{radford2021learningtransferablevisualmodels}
        };


        \node[below, text width=3cm, align=center,text=Red] at (19,2.2){\textbf{VQ-Diffusion}\cite{gu2022vectorquantizeddiffusionmodel}};
        
        \node[below, text width=3cm, align=center,text=Red] at (19,1.7)
        {\textbf{DALLE-2}\cite{ramesh2022hierarchicaltextconditionalimagegeneration}
        };

        \node[below, text width=3cm, align=center,text=Green] at (19,2.7)
        {\textbf{DF-GAN}\cite{tao2022dfgansimpleeffectivebaseline}
        };

        \node[below, text width=3cm, align=center,text=Yellow] at (19,3.2)
        {\textbf{VQGAN-CLIP}\cite{crowson2022vqganclipopendomainimage}
        };

        \node[below, text width=3cm, align=center,text=Red] at (19,4.2)
        {\textbf{Stable Diffusion}\cite{rombach2022highresolutionimagesynthesislatent}
        };

        \node[below, text width=3cm, align=center,text=Pink] at (19,5.2)
        {\textbf{Flow Matching}\cite{lipman2023flowmatchinggenerativemodeling}
        };


        \node[below, text width=3cm, align=center,text=Red] at (21,-1.2){\textbf{GILL}\cite{koh2023generatingimagesmultimodallanguage}};
        
        \node[below, text width=3cm, align=center,text=Green] at (21,-1.7)
        {\textbf{GALIP}\cite{tao2023galipgenerativeadversarialclips}
        };

        \node[below, text width=3cm, align=center,text=Green] at (21,-2.2)
        {\textbf{GigaGAN}\cite{kang2023gigagan}
        };

        \node[below, text width=3cm, align=center,text=Yellow] at (21,-2.7){\textbf{DALLE-3}\cite{BetkerImprovingIG}
        };

        \node[below, text width=3cm, align=center,text=Red] at (21,-3.2){\textbf{DreamBooth}\cite{ruiz2023dreamboothfinetuningtexttoimage}
        };

        \node[below, text width=3cm, align=center,text=Red] at (21,-3.7){\textbf{ControlNet}\cite{gu2024jiehuapaintingsstylefeature}
        };

        \node[below, text width=3cm, align=center,text=Red] at (21,-4.2){\textbf{Inst}\cite{Zhang_2023_inst}
        };

        \node[below, text width=3cm, align=center,text=Red] at (21,-4.7){\textbf{ControlNet}\cite{zhang2023addingconditionalcontroltextto}
        };

        \node[below, text width=3cm, align=center,text=Red] at (23,2.2){\textbf{Stable Diffusion 3}\cite{esser2024scalingrectifiedflowtransformers}};
        \node[below, text width=3cm, align=center,text=Red] at (23,2.7)
        {\textbf{Style Injection}\cite{chung2024style}
        };
        \node[below, text width=3cm, align=center,text=Red] at (23,3.2)
        {\textbf{DiffStyler}\cite{li2024diffstyler}
        };
        \node[below, text width=3cm, align=center,text=Blue] at (23,3.7){\textbf{VAR}\cite{tian2024visualautoregressivemodelingscalable}
        };
        \node[below, text width=3cm, align=center,text=Blue] at (23,4.2){\textbf{RandAR}\cite{pang2024randardecoderonlyautoregressivevisual}
        };
        \node[below, text width=3cm, align=center,text=Red] at (23,4.7)
        {\textbf{Sana}\cite{xie2024sana}
        };
        \node[below, text width=3cm, align=center,text=Blue] at (23,5.2)
        {\textbf{Infinity}\cite{Infinity2024}
        };
        \node[below, text width=3cm, align=center,text=Blue] at (23,5.7)
        {\textbf{MAR}\cite{li2024autoregressiveimagegenerationvector}
        };

        \node[above, text width=3cm, align=center,text=Blue] at (25,-1.8) {\textbf{EditAR}\cite{mu2025editarunifiedconditionalgeneration}
        };
        \node[below, text width=3cm, align=center,text=Red] at (25,-2.0)
        {\textbf{SDXL-Turbo}
        };

         \node[below, text width=3cm, align=center,text=Pink] at (25,-2.5)
        {\textbf{Diff2Flow}\cite{schusterbauer2025diff2flowtrainingflowmatching}
        };
        \node[below, text width=3cm, align=center,text=Blue] at (25,-3.0)
        {\textbf{DAR}\cite{xu2025directionaware}
        };

    \end{tikzpicture}
    \caption{
    Timeline of key milestones and innovations in style transfer and generative models. The \textcolor{Yellow}{traditional methods}, \textcolor{Purple}{VAE-based methods}, \textcolor{Green}{GAN-based methods}, \textcolor{Pink}{Flow matching}, \textcolor{Blue}{Autoregressive} and \textcolor{Red}{diffusion-based methods} are highlighted in \textcolor{Yellow}{yellow}, \textcolor{Purple}{purple}, \textcolor{Green}{green}, \textcolor{Pink}{pink}, \textcolor{Blue}{blue} and \textcolor{Red}{red}, respectively.
}
    \label{fig:style_timeline}
\end{figure*}
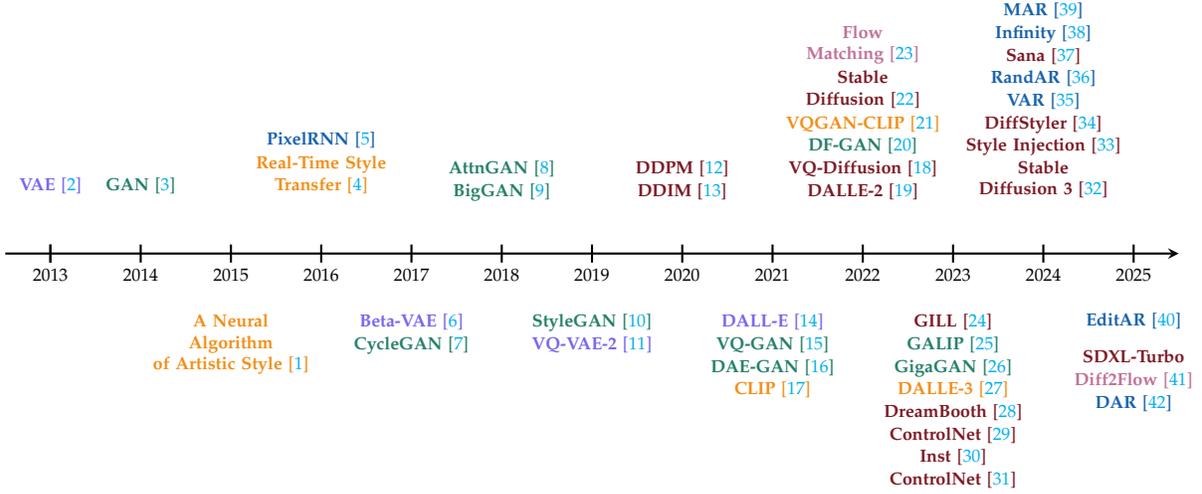

\subsection{Technological Evolution}
Early computer graphics heuristics evolved into deep generative modelling
after the advent of Variational Autoencoders\cite{kingma2013auto} and Generative Adversarial Networks\cite{goodfellow2014generative}. The landmark Neural Style Transfer of Gatys\cite{gatys2015neural} revealed that Gram statistics encode visual style, spawning real-time feed-forward variants and arbitrary style transfer. Diffusion models\cite{ho2020denoising} and large-scale text–image systems(Stable Diffusion\cite{rombach2022highresolutionimagesynthesislatent} and DALL-E2) pushed resolution and semantic control further, while CLIP\cite{radford2021learning} enabled language-conditioned stylisation(StyleGAN-NADA\cite{gal2021stylegannada}, CLIPstyler\cite{kwon2022clipstylerimagestyletransfer}). Recent work explores structural decomposition and efficient architectures that achieve real-time 4K stylisation with tight content preservation. Figure \ref{fig:style_timeline} summarises these milestones.

\subsection{Research Objectives and Contributions}
This review examines recent advances in style transfer, with a focus on generative models.

\begin{enumerate}
    \item Generative Models:  Survey of GAN, VAE, autoregressive, and diffusion style approaches, focusing on image quality, training stability, and style–content control.  
    \item Evaluation Frameworks:  Discussion of effectiveness, robustness, controllability, interpretability, practicality, and aesthetic quality across metrics.
    \item Applications and Future Trends:  Summary of current use-cases (portrait, video, text, 3-D, etc.) and likely research directions.  
    \item Datasets and Metrics:  Collation of publicly available datasets and benchmark criteria relevant to style-transfer research.  
\end{enumerate}

This paper is structured as follows: Section 2 covers the foundations of style transfer, Section 3 explores generative models, Section 4 discusses evaluation frameworks, Section 5 examines domain-specific applications, and Section 6 reviews datasets and evaluation methodologies. Additional resources are available at \url{https://github.com/neptune-T/Awesome-Style-Transfer}.

\section{Fundamental Theories of Style Transfer}
\subsection{Separation and Representation of Style and Content}

Neural style transfer decomposes an image into two independent factors.  The \emph{content} component—object geometry and spatial layout—is described by deep‐layer activations \(\Phi_\ell(I)\) of a fixed CNN, so that \(\Phi_\ell(I_c)\) and \(\Phi_\ell(I_t)\) should remain close to preserve structure.  

\begin{figure}[h]
  \centering
  \includegraphics[width=\linewidth]{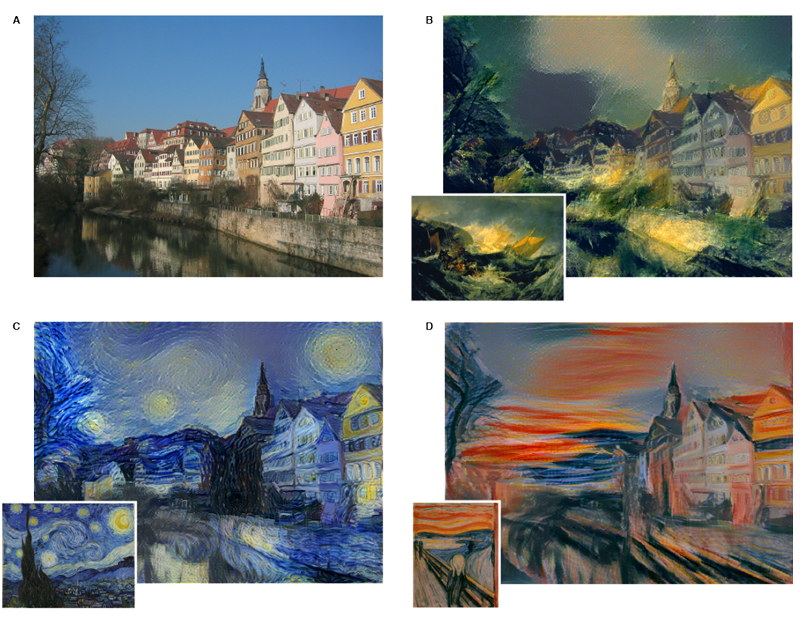}
  \caption{Neural-style-transfer result reproduced from Gatys \textit{et al.}\ \cite{gatys2015neural}.}
  \label{fig:NST_example}
\end{figure}

As illustrated in Figure~\ref{fig:NST_example}, NST synthesizes content and style from separate images to generate a single output, preserving structural integrity while incorporating the artistic characteristics of the style image. The figure demonstrates the transformation of a photograph into artworks inspired by famous paintings.

The \emph{style} component, in turn, is characterised by statistics of shallower features.  
Gatys \textit{et al.} first captured appearance with the Gram matrix
\begin{equation}
  G_\ell(I)=\Phi_\ell(I)\,\Phi_\ell(I)^{\!\top},
\end{equation}

whose entries measure second-order channel correlations and thus encode global colour‐texture patterns.  
Subsequent work showed that real-time, arbitrary stylisation can be achieved by matching only the first two channel moments.  Given content features \(x_c\) and style features \(x_s\) at layer \(\ell\), Adaptive Instance Normalisation (AdaIN)\cite{huang2017arbitrarystyletransferrealtime} applies the affine transform
\begin{equation}
    \operatorname{AdaIN}(x_c,x_s)=
  \sigma(x_s)\,
  \frac{x_c-\mu(x_c)}{\sigma(x_c)}
  +\mu(x_s),
\end{equation}

where \(\mu(\cdot)\) and \(\sigma(\cdot)\) denote per-channel mean and standard deviation. More recently, language–image encoders such as CLIP\cite{radford2021learningtransferablevisualmodels} provide a cross-modal alternative, allowing style cues to be expressed directly as text prompts.

These representations give rise to the Gram- and AdaIN-based losses detailed in \ref{loss}.

\subsection{Loss Functions for Content–Style Disentanglement}\label{loss}

Let \(I_c\), \(I_s\), and \(I_t\) denote the content, style, and generated images, respectively.

\textbf{Content loss.}  
\begin{equation}
  \mathcal{L}_{\text{content}}
  =
  \sum_{\ell\in\mathcal{C}}
  \tfrac12
  \bigl\lVert\,
    \Phi_\ell(I_t)-\Phi_\ell(I_c)
  \bigr\rVert_2^{2},
\end{equation}
where \(\mathcal{C}\) indexes geometry‐sensitive layers.

\textbf{Style loss (Gram).}  
\begin{equation}
  \mathcal{L}_{\text{style}}
  =
  \sum_{\ell\in\mathcal{S}}
  \frac{1}{4C_\ell^{2}H_\ell^{2}W_\ell^{2}}
  \bigl\lVert
    G_\ell(I_t)-G_\ell(I_s)
  \bigr\rVert_2^{2},
\end{equation}
with \(C_\ell\) channels.

\textbf{Style loss (AdaIN).}  
\begin{equation}
  \mathcal{L}_{\text{style}}^{\text{AdaIN}}
  =
  \sum_{\ell\in\mathcal{S}}
  \Bigl(
    \bigl\lVert\mu_t^\ell-\mu_s^\ell\bigr\rVert_2^{2}
    +
    \bigl\lVert\sigma_t^\ell-\sigma_s^\ell\bigr\rVert_2^{2}
  \Bigr),
  \label{eq:adain_loss}
\end{equation}
where \(\mu^\ell\) and \(\sigma^\ell\) are the per–channel mean and standard deviation.

\textbf{Total objective.}  
\begin{equation}
  \mathcal{L}(I_t)=
  \alpha\,\mathcal{L}_{\text{content}}
  +\beta\,\mathcal{L}_{\text{style}}
  +\gamma\,\mathcal{L}_{\text{TV}},
  \label{eq:overall_loss}
\end{equation}
\(\alpha,\beta,\gamma>0\) are user weights and \(\mathcal{L}_{\text{TV}}\) is a total-variation term.
Image \(I_t\) is obtained by gradient descent on Eq.~\eqref{eq:overall_loss}.
In practice, replacing the Gram term with \(\mathcal{L}_{\text{style}}^{\text{AdaIN}}\) trades the
\(\mathcal{O}(C^{2})\) complexity of Gram matrices for \(\mathcal{O}(C)\) while remaining perceptually faithful.

During optimisation, we compute $\Phi_\ell(\cdot)$ with a fixed VGG-19 encoder. For the Gram variant, $\mathcal{L}_{\text{style}}$ in Eq.~\eqref{eq:overall_loss}
is set to the correlation loss above; for the AdaIN variant we
\emph{replace} that term by $\mathcal{L}_{\text{style}}^{\text{AdaIN}}$
(Equation~\ref{eq:adain_loss}) and—following~\cite{huang2017arbitrarystyletransferrealtime}—
apply the AdaIN affine transform to the \emph{decoder} feature maps
at every forward pass.
The remainder of the training pipeline (learning rate, TV weight, optimiser)
is identical, so switching between Gram and AdaIN requires only
changing this single loss component.

\subsection{Image Processing and Hand-Crafted Style-Transfer Techniques}
From Gram-based neural style transfer of Gatys \textit{et al.} \cite{gatys2015neural} 
to feed-forward fast stylisation \cite{johnson2016perceptual,ulyanov2016texture} 
and AdaIN-based arbitrary transfer \cite{huang2017arbitrarystyletransferrealtime}, 
research has progressively traded optimisation time for flexibility. 
Recent work extends control (SEAN \cite{zhu2020sean}), data efficiency (DIP \cite{ulyanov2018deep}), 
and resolution (StyTr2 \cite{deng2022stytr2}); a concise comparison is given in Table \ref{tab:style_transfer_methods}.

\begin{table*}[h!]
\centering
\caption{Chronologically Organized Overview of Style Transfer Methods.}
\label{tab:style_transfer_methods}
\begin{tblr}{
  colspec = {clll},
  row{1} = {bg=gray!30, font=\bfseries},
  row{3-5} = {bg=gray!10},
  row{11-13} = {bg=gray!10},
  row{15} = {bg=gray!10},
  row{19} = {bg=gray!10},
  row{23-29} = {bg=gray!10},
  hline{2,3,6,11,14,15,16,19,20,23,30} = {1-4}{gray!60},
}
Year & Method & Publication & Innovation \\
2015 & NST \cite{gatys2015neural} & CVPR & CNN-based method separates and recombines content and style features \\

~ & Fast Transfer \cite{johnson2016perceptual} & ECCV & Uses perceptual loss for real-time style transfer with high quality \\
    2016 & Texture Net \cite{ulyanov2016texture} & arXiv & Introduces fast texture-based network optimized for style transfer \\
    ~ & Instance Norm \cite{ulyanov2016instance} & arXiv & Presents Instance Normalization to enhance style transfer by normalizing features \\
 ~ & Universal Style \cite{li2017universal} & NIPS & Develops universal approach for transferring any style without retraining \\
    ~ & Light Transfer \cite{ulyanov2017improved} & CVPR & Proposes lightweight network enabling real-time transfer on mobile devices \\
    2017 & AdaIN \cite{huang2017arbitrary} & ICCV & Introduces Adaptive Instance Normalization to align content and style features \\
    ~ & Histogram Loss \cite{risser2017histogram} & CVPR & Employs histogram matching to ensure color consistency between images \\
    ~ & Deep Photo \cite{Luan2017DeepPS} & CVPR & Applies affine color transforms to preserve structure and realism\\
    
    ~ & Avatar-Net \cite{sheng2018avatarnetmultiscalezeroshotstyle} & CVPR & Develops multi-scale zero-shot framework for style transfer without paired data \\
    2018 & MUNIT \cite{huang2018multimodal} & CVPR & Extends AdaIN for multimodal transfer, allowing diverse style outputs \\
    ~ & NST-MetaNet \cite{shen2018style} & CVPR & Introduces meta-learning for quick adaptation to new styles in style transfer \\
    
2019 & AMST \cite{yao2019attentionawaremultistrokestyletransfer} & CVPR & Attention-aware multi-stroke style transfer for detailed artistic styles \\
2020 & SEAN \cite{zhu2020sean} & CVPR & Implements semantic-based multi-region style control for distinct image regions \\
~ & AdaAttN \cite{liu2021adaattnrevisitattentionmechanism} & ICCV & Combines adaptive normalization with attention for precise style control \\
    2021 & ArtFlow \cite{an2021artflowunbiasedimagestyle} & CVPR & Achieves lossless transfer by preserving content structure while applying styles \\
    ~ & CLIPstyler \cite{kwon2022clipstylerimagestyletransfer} & CVPR & Utilizes CLIP for text-guided style transfer via textual descriptions \\
2022 & CSD-AST \cite{kotovenko2019iccv} & ICCV & Cross-spectral domain style transfer handling different frequency components \\

~ & Composer\cite{Huang2023ComposerCA} & ICML & Enables controllable image synthesis with composable conditions \\
2023 & RCST \cite{kang2023regioncontrolledstyletransfer} & arXiv & Region-controlled style transfer for selective style application \\
    ~ & OIT-SD \cite{huang2023optimalimagetransportsparse} & arXiv & Optimal Image Transport on Sparse Dictionaries for efficient feature transport \\
~ & Puff-Net \cite{zheng2024puffnetefficientstyletransfer} & CVPR & Lightweight style transfer with efficient feature extraction and blending \\
    ~ & S2WAT \cite{zhang2023s2watimagestyletransfer} & AAAI & Style-to-Weather Adaptation Transfer for context-aware style application \\
    ~ & EUC-HSTN \cite{KHOWAJA2024e27012} & Heliyon & Enhanced Universal Composition with Hybrid Style Transfer Network \\
    2024 & ReLU-Oscillator \cite{SHI2024124510} & ESWA & Introduces oscillatory behavior in activations for pattern generation in style transfer \\
    ~ & AEANet \cite{li2024aeanetaffinityenhancedattentional} & arXiv & Affinity-enhanced attentional network for improved feature matching in style transfer \\
    ~ & ANST-AS \cite{li2024artisticneuralstyletransfer} & arXiv & Artistic Neural Style Transfer with Activation Smoothing for better style blending \\
    ~ & StyleMamba \cite{wang2024stylemambastatespace} & arXiv & State-space model for dynamic, temporal style transfer in videos \\
\end{tblr}
\end{table*}

\subsection{Variational Autoencoder (VAE)}

A Variational Autoencoder (VAE) \cite{kingma2013auto} couples an \emph{encoder}
$q_{\boldsymbol\phi}(z\,|\,x)$ with a \emph{decoder} $p_{\boldsymbol\theta}(x\,|\,z)$
to learn a continuous latent manifold.
The encoder maps an input image \(x\) to a Gaussian posterior over
latent code \(z\), while the decoder reconstructs \(x\) from samples of that
posterior.
Training maximises the evidence lower bound (ELBO)

\[
\mathcal{L}_{\text{VAE}}(x;\boldsymbol\theta,\boldsymbol\phi)=
\underbrace{\mathbb{E}_{q_{\boldsymbol\phi}(z|x)}
            \bigl[\log p_{\boldsymbol\theta}(x\,|\,z)\bigr]}_{\text{reconstruction}}
-
\underbrace{D_{\mathrm{KL}}\!\bigl(q_{\boldsymbol\phi}(z|x)\,\|\,p(z)\bigr)}_{\text{regularisation}},
\]

where the prior is fixed to \(p(z)=\mathcal{N}(0,I)\).
The first term enforces pixel-level fidelity,
and the KL term pulls the approximate posterior towards the prior, yielding a smooth latent space that supports interpolation and random sampling.
After training, new images are generated by drawing \(z\!\sim\!p(z)\)
and decoding through \(p_{\boldsymbol\theta}(x\,|\,z)\).

\subsection{Generative Adversarial Network (GAN)}

A Generative Adversarial Network(GAN) \cite{goodfellow2014generative} comprises two neural nets trained in opposition. The \emph{generator} \(G_{\boldsymbol\theta}\) transforms latent noise
\(z\!\sim\!p_z(z)=\mathcal N(0,I)\) into synthetic samples \(G(z)\); the \emph{discriminator} \(D_{\boldsymbol\psi}\) scores whether a sample comes from the data distribution \(p_{\text{data}}(x)\) or from \(G\). The game‐theoretic objective is:

\begin{equation}
\begin{aligned}
\min_{\theta}\max_{\psi}\,\mathcal V(D,G)
  &= \mathbb{E}_{x\sim p_{\text{data}}}\!\bigl[\log D_\psi(x)\bigr] \\
  &\;+\mathbb{E}_{z\sim p_z}\!\bigl[\log\!\bigl(1-D_\psi(G_\theta(z))\bigr)\bigr].
\end{aligned}
\end{equation}

At the Nash equilibrium the generator distribution \(p_g\) matches\(p_{\text{data}}\), minimising the Jensen–Shannon divergence between them.

\subsection{Flow-Matching Models}

Flow-matching (FM) \cite{lipman2023flowmatchinggenerativemodeling} frames generation
as learning a time-dependent velocity field that deterministically transports
Gaussian noise to data.
Given $x_0\!\sim\!p_{\text{data}}$ and
$
  x_t=\sqrt{1-t}\,x_0+\sqrt{t}\,\varepsilon,\;
  \varepsilon\!\sim\!\mathcal N(0,I),\;
  t\!\sim\!\mathcal U(0,1),
$
the target flow is
$u_t(x_t)=\tfrac12(\varepsilon-x_0)$.
A neural field $f_\theta$ is fitted via

\begin{equation}
\label{eq:fm}
\mathcal{L}_{\text{FM}}
=\mathbb E_{t,x_0,\varepsilon}
  \bigl\|
    f_\theta(x_t,t)-u_t(x_t)
  \bigr\|_2^{2}.
\end{equation}

At inference, integrating the ODE
$\dot x_t=f_\theta(x_t,t)$ from $t{=}1\!\to\!0$
generates a sample in tens of steps—significantly faster than
classic diffusion samplers.

\subsection{Autoregressive (AR) Models}

Autoregressive models factorise the joint likelihood of an image
into a product of conditionals over pixels or discrete latent tokens:
\begin{equation}
    p(x) = \prod_{i=1}^{N} p(x_i \mid x_1, x_2, ..., x_{i-1}),
\label{eq_ar}
\end{equation}
where \(x_i\) is the \(i\)-th pixel (or token) and
\(x_{<i}\) denotes all previously generated elements.
Sequential decoding allows exact likelihood evaluation and
fine-grained structural control, which is exploited in text-to-image
pipelines (e.g.\ DALL·E \cite{ramesh2021zeroshottexttoimagegeneration}, VAR \cite{tian2024visualautoregressivemodelingscalable}) and layout-conditioned style transfer.

\subsection{Diffusion Models}

Diffusion models \cite{sohl2015deep,ho2020denoising} generate images by reversing a Markovian noise process.
The \emph{forward} (diffusion) process gradually corrupts a data sample
\(x_0\!\sim\!p_{\text{data}}\) into Gaussian noise \(x_T\) through

\begin{equation}
\label{eq:fwd_diff}
q(x_t\!\mid\!x_{t-1})=
\mathcal{N}\!\bigl(
  x_t\,;\,
  \sqrt{1-\beta_t}\,x_{t-1},\,
  \beta_t I
\bigr),\qquad t=1\ldots T,
\end{equation}
with a variance schedule \(\{\beta_t\}_{t=1}^{T}\).

The \emph{reverse} (generative) process learns a neural approximation
\(p_\theta(x_{t-1}\!\mid\!x_t)\) of the intractable backward transition:
\begin{equation}
\label{eq:rev_diff}
p_\theta(x_{t-1}\!\mid\!x_t)=
\mathcal{N}\!\bigl(
  x_{t-1}\,;\,
  \mu_\theta(x_t,t),\,
  \Sigma_\theta(x_t,t)
\bigr),
\end{equation}
trained by minimising the variational lower bound,
which is equivalent to a weighted denoising score-matching objective.
Sampling is performed by iteratively applying \eqref{eq:rev_diff}
from \(t=T\) down to \(0\), or more efficiently via
ODE/flow solvers that cut the step count to a few dozen while retaining image fidelity.

\section{Development and Application of Generative Models in Style Transfer}

\subsection{VAE-Based Style Transfer}
As shown in Table \ref{tab:vae_methods}, the application of VAE in style transfer has gradually evolved from the basic model to various improved methods. In this survey, we discuss the application of VAE in style transfer in three main areas: \textbf{improvements in latent space structure}  \cite{choi2018hierarchical,zhang2020conditional,kingma2018glow,dinh2017real,rezende2015variational,huang2021conditional}, \textbf{integration with adversarial generative models}\cite{larsen2015autoencoding,yang2018vae,tolstikhin2017wasserstein,wang2020paintgan}, \textbf{efficient implementations of disentangled learning in style transfer} \cite{higgins2017beta,choi2018multi,chen2020adaptive,oord2017neural,liu2020conditional}. The technological innovations and progress in each area are detailed in the Appendix, where the readers can explore the specific methods and technical background.

These continuously evolving VAE methods have not only improved the quality of style transfer but also expanded its range of applications, driving rapid advances in fields such as artistic style transfer and image-to-image translation. The subsequent sections will provide a detailed introduction to the historical evolution, technical details, and specific applications of these methods in style transfer, with further information available in the Appendix \ref{vae_add}.

\begin{table*}[h!]
\caption{Summary of Key VAE Variants and Their Innovations.}
\label{tab:vae_methods}
\begin{tblr}{
  colspec = {clll},
  row{1} = {bg=gray!30, font=\bfseries},
  row{3} = {bg=gray!10},
  row{5-9} = {bg=gray!10},
  row{13-15} = {bg=gray!10},
  row{19-21} = {bg=gray!10},
  row{25-27} = {bg=gray!10},
  row{29} = {bg=gray!10},
  hline{2,3,4,5,10,13,16,19,22,25,28,29,30} = {1-4}{gray!60},
}
Year & Method & Publication & Innovation \\

2013 & VAE \cite{kingma2013auto}  & ICLR & Variational inference for generative models  \\

2015 & Conditional VAE \cite{sohn2015learning}  & NeurIPS & Conditioning on attributes for controlled generation  \\

2016 & Hierarchical VAE \cite{sonderby2016ladder}  & NeurIPS & Multi-scale latent variable modeling \\

~ & $\beta$-VAE \cite{higgins2017beta} & ICLR & Disentangled representation learning with KL term scaling  \\
~ & VQ-VAE \cite{van2017neural} & NeurIPS & Discrete latent representation through vector quantization  \\
2017 & InfoVAE \cite{zhao2017info} & arXiv & MMD penalty instead of KL divergence for flexible priors \\
~ & WAE \cite{tolstikhin2019wassersteinautoencoders} &  ICLR & WAE minimizes Wasserstein distance for improved latent representation  \\
~ & MSVAE\cite{Cai2017MultiStageVA} & SDM & Hierarchical multi-stage framework for coarse-to-fine image generation \\
~ & FactorVAE \cite{kim2018disentangling}  & ICML & Total correlation minimization for improved disentanglement \\
2018 & DIP-VAE \cite{kumar2018variational}  & ICLR & Disentanglement via covariance regularization  \\
~ & $\beta$-TCVAE \cite{chen2018isolating}  & ICML & Explicit total correlation term to reduce redundancy  \\
~ & BigVAE \cite{child2019biggan}  & ICLR & Large-scale VAE with hierarchical latent spaces  \\
2019 & VQ-VAE-2 \cite{razavi2019generating}  & NeurIPS & Hierarchical VQ-VAE with multi-stage encoding  \\
~ & CVAE-ESF \cite{Abid2019ContrastiveVA} & arXiv & Integrates contrastive learning with VAEs to enhance salient feature representation \\
~ & NVAE \cite{vahdat2020nvae} & NeurIPS & Efficient hierarchical VAE with normalizing flows  \\
2020 & Parallel VAE \cite{oord2018parallel}  & ICML & Parallel training for faster VAE convergence  \\
~ & DGIR-IGF\cite{Ma2021DecouplingGA} & ICLR & Decouples global and local representations using invertible generative flows \\
~ & StyleVAE \cite{xu2021variational}  & CVPR & Integrating style transfer into VAE framework \\
2021 & Score-VAE \cite{vahdat2021score}  & NeurIPS & Using score matching with VAE for sharper generation \\
~ & Shape your Space\cite{NEURIPS2021_3c057cb2} & NeurIPS & Introduces Gaussian Mixture regularization to deterministic autoencoders \\
~ & StyleMeUp\cite{Sain_2021_CVPR} & CVPR & Proposes a style-agnostic framework for sketch-based image retrieval \\
2022 & Distilled-VAE \cite{shen2022distillation}  & ICML & Distilling knowledge into compact VAE  \\
~ & GRS-FSC\cite{Xu2022GeneratingRS} & CVPR & Generates high-quality representative samples to improve few-shot classification \\

~ & Masked-VAE \cite{bao2023masked} & CVPR & Masked modeling for VAE to enhance detail  \\
2023 & DynamicVAE \cite{gao2023dynamic}  & AAAI & Adaptive latent space for dynamic content  \\
~ & Catch Missing Details \cite{inproceedings}& CVPR & Augments VAEs with frequency domain information to capture missing details in images \\

2024 & RFE-DiffSketch \cite{yun2024representativefeatureextractiondiffusion} &  arXiv & Feature extraction during diffusion process combined with VAE \\

2025 & REPA-E \cite{leng2025repaeunlockingvaeendtoend} & arXiv & uses representation‑alignment loss for end‑to‑end VAE‑diffusion training \\ 
\end{tblr}
\centering
\end{table*}

\subsection{GAN-Based Style Transfer}

Generative Adversarial Networks (GANs) have become one of the most powerful tools for style transfer, thanks to their ability to learn and generate realistic images through an adversarial framework. Despite their effectiveness, GANs still face challenges in style transfer tasks, such as maintaining content consistency, handling complex styles, and improving training stability. Over time, innovations in \textbf{latent space modeling} \cite{zhu2017unpaired, karras2019style, karras2019stylegan2},\cite{choi2018stargan} \textbf{training stability}\cite{arjovsky2017wasserstein,brock2019largescalegantraining,larsen2015autoencoding,denton2015deepgenerativeimagemodels,jolicoeurmartineau2018relativisticdiscriminatorkeyelement} and \textbf{disentanglement learning} \cite{higgins2017beta,zhang2020deformgananunsupervisedlearningmodel,odena2017conditionalimagesynthesisauxiliary,mirza2014conditionalgenerativeadversarialnets} have significantly improved the performance of GAN-based models in this domain.

Given space constraints, this section provides only a brief mention of GAN-based approaches, while a more detailed discussion is available in Appendix \ref{gan_add}. Table \ref{tab:gan_variants} offers a reference for understanding their progression and diversification in style transfer.

\begin{table*}
\caption{Summary of Key GAN Variants and Their Innovations.}
\label{tab:gan_variants}
\DefTblrTemplate{contfoot-text}{default}{}
\DefTblrTemplate{conthead-text}{default}{}
\DefTblrTemplate{caption}{default}{}
\DefTblrTemplate{conthead}{default}{}
\DefTblrTemplate{capcont}{default}{}
\SetTblrTemplate{contfoot}{}
\SetTblrTemplate{conthead}{}
\begin{tblr}{
  colspec = {clll},
  row{1} = {bg=gray!30, font=\bfseries},
  row{3} = {bg=gray!10},
  row{5-17} = {bg=gray!10},
  row{25-27} = {bg=gray!10},
  row{35-37} = {bg=gray!10},
  row{43-49} = {bg=gray!10},
  row{53} = {bg=gray!10},
  hline{2,3,4,5,18,25,28,35,38,43,50,53} = {1-4}{gray!60},
}
Year & Method & Publication & Innovation \\
2014 & GAN \cite{goodfellow2014generative}  & NeurIPS & Introduced adversarial training for generative models \\
2015 & DCGAN \cite{radford2015unsupervised}  & ICLR &Incorporated CNN layers for stable training and improved image quality \\ 
2016 & ImprovedGAN \cite{salimans2016improvedgan}  & NeurIPS & Techniques for training stability and diversity in GANs \\
~ & VideoGAN \cite{vondrick2016generating}  & NeurIPS & GANs for realistic video generation \\
~ & InfoGAN \cite{chen2016infogan}  & NeurIPS & Introduced disentanglement by maximizing mutual information \\
~ & CycleGAN \cite{zhu2017unpaired}  & ICCV & Enabled unpaired image-to-image translation with cycle-consistency loss \\
~ & PGGAN \cite{karras2017progressive}  & ICLR & Progressive growing method for high-resolution image synthesis \\
~ & WGAN \cite{arjovsky2017wasserstein}  & ICML & Introduced Wasserstein distance to address mode collapse \\
~ & LSGAN \cite{mao2017least}  & ICCV & Used least-squares loss for smoother optimization \\
2017 & BEGAN \cite{berthelot2017began}  & ICLR & Balanced generator and discriminator training with autoencoder loss \\
~ & Pix2Pix \cite{isola2017pix2pix}  & CVPR & Conditional GAN for paired image-to-image translation \\
~ & FID-GAN \cite{heusel2017gans}  & NeurIPS & Enhanced evaluation metric for GANs based on Fréchet Inception Distance \\
~ & DualGAN \cite{yi2017dualgan}  & ICCV & Bidirectional image-to-image translation for paired domains \\
~ & SeqGAN \cite{yu2017seqgan}  & AAAI & Sequence generation for text via GAN framework \\
~ & SRGAN \cite{ledig2017srgan}  & CVPR & Super-resolution GAN for high-quality upscaling \\
~ & ComboGAN\cite{anoosheh2017comboganunrestrainedscalabilityimage} & CVPR & Scalable multi-domain image translation \\
~   & PoseNormGAN\cite{qian2018posenormalizedimagegenerationperson} & ECCV & Pose-normalized person re-identification \\
~  & DG-GAN \cite{tang2019dualgeneratorgenerativeadversarial} & ACCV & Dual generators disentangle multi-domain mappings \\
~ & AsymmetricGAN\cite{tang2019asymmetric} & ACCV & Asymmetric GANs decouple encoder-decoder complexities \\
2018 & BigGAN \cite{brock2018large}  & ICLR & Large-scale GAN trained on ImageNet for high-resolution synthesis \\
~  & GANimation \cite{pumarola2018ganimation}  & ECCV & GAN for facial animation with emotional expressions \\
~  & StarGAN \cite{choi2018stargan}  & CVPR & Multi-domain image-to-image translation within a single model \\
~  & MaskGAN\cite{fedus2018maskganbettertextgeneration} & ICLR & Text generation via GAN-based in-filling \\
~ & GauGAN \cite{park2019gaugan}  & CVPR & Semantic image synthesis with user-guided input \\
2019   & StyleGAN \cite{karras2019style}  & CVPR & Style-based generator architecture with control over image features \\
~  & SAGAN \cite{zhang2019sagan}  & ICML & Self-attention GAN for better long-range dependencies in images \\
~ & StyleGAN2 \cite{karras2020analyzing} & CVPR & Improved style control and synthesis quality \\
~   & ConSinGAN \cite{hinz2020consingan}  & CVPR & Single-image GAN for diverse content generation \\
~   & PaintGAN \cite{wang2020paintgan}  & ECCV & Brushstroke control for artistic style transfer \\
2020  & AD-GAN\cite{men2020controllablepersonimagesynthesis} & CVPR & Generates person images with controllable attributes using decomposed GAN \\
~ & LGGAN\cite{tang2020local2} & CVPR &
Combines class-specific and global cues for semantic generation \\
~  & Old Photo Restoration\cite{wan2020oldphotorestorationdeep} & CVPR & Restores old photos using triplet domain translation and VAEs \\
~  & NICE-GAN\cite{chen2020reusingdiscriminatorsencodingunsupervised} & CVPR & Reuses GAN discriminators as encoders for unsupervised image translation \\
~ & StyleGAN3 \cite{karras2021stylegan3}  & NeurIPS & Mitigated aliasing issues for improved consistency \\
  2021  & GANformer \cite{hudson2021ganformer}  & CVPR & Hybrid attention GAN combining transformers and GANs \\
~ & CoModGAN \cite{sauer2021comodgan}  & CVPR & Conditional modulation for high-resolution inpainting \\
~ & StyleSwin \cite{zhang2022styleswin}  & CVPR & Transformer-based style synthesis for flexible architecture \\
~  & InstGAN \cite{wang2022instgan}  & CVPR & Instance-based GAN for diverse object generation \\
2022   & style-aware-discriminator\cite{kim2022styleawarediscriminatorcontrollableimage} & CVPR & Integrates style encoding into discriminator for controllable image translation \\
~ & LGGAN\cite{tang2022local} & CVPR &
Semantic-aware upsampling bridges local-global feature generation \\
~ & DF-GAN \cite{tao2022df} & CVPR & One-stage GAN with deep fusion and target-aware discrimination \\
~ & EGCL-GAN\cite{tang2023edge} & ICLR & Edge guidance and contrastive loss enhance structure-aware synthesis \\
~ & PI-TRANS\cite{ren2023pitransparallelconvmlpimplicittransformationbased} & ICASSP & ConvMLP and implicit warping enable efficient view translation \\  
~  & GigaGAN\cite{kang2023gigagan} & CVPR & High-resolution image synthesis with hierarchical multi-scale generation \\
2023  & DeltaEdit\cite{lyu2023deltaedit} & CVPR & Precise semantic image editing through delta feature manipulation \\
~  & NoisyTwins\cite{rangwani2023noisytwins} & CVPR & Self-supervised learning with noisy twin networks for robust generation \\
~ & CREPS\cite{thuan2023creps} & CVPR & Cross-region event propagation for semantic image synthesis \\
~ & GALIP\cite{tao2023galip} & CVPR & CLIP-guided GAN aligns text-image semantics effectively \\
~ & PaGoDA \cite{kim2024pagodaprogressivegrowingonestep} & NeurIPS & progressively grows a single‑step generator to upscale diffusion models \\
2024  & CCST-GAN\cite{10586662} & ICIP & Content-conditioned style transfer with adaptive normalization \\ 
~ & SiD\cite{zhou2024adversarialscoreidentitydistillation} &  ICLR & enhances SiD by adding adversarial loss for one-step distillation \\
2025 & DDO \cite{zheng2025direct} & ICML & implicit GAN discriminator finetunes diffusion models via likelihood‑ratio \\
\end{tblr}
\end{table*}

\subsection{Diffusion-Based Style Transfer}

Diffusion models have emerged as a transformative framework in generative modeling, demonstrating remarkable capabilities in image synthesis and style transfer. This section explores the advancements in diffusion-based techniques, focusing on four key areas: \textbf{Enhanced Latent Representations for Style Transfer} \cite{song2020generativemodelingestimatinggradients,dhariwal2021diffusion,mardia2016scorematchingestimatorsdirectional}, \textbf{Optimized Training Processes for Stability and Scalability} \cite{yang2024consistencyflowmatchingdefining,zhang2023gddimgeneralizeddenoisingdiffusion,NEURIPS2021_958c5305,NEURIPS2021_b578f2a5,yang2024crossmodal}, \textbf{Advanced Conditioning Mechanisms for Multi-Modal and Cross-Domain Style Transfer} \cite{ramesh2021zeroshottexttoimagegeneration,zhang2023prospectpromptspectrumattributeaware,Kim_2022_CVPR,zuo2024multiviewconsistentstyletransfer,hachnochi2023crossdomaincompositingpretraineddiffusion,saharia2022photorealistic,ramesh2022hierarchicaltextconditionalimagegeneration}, and \textbf{Disentanglement and Content Preservation in Diffusion Models} \cite{wang2023stylediffusioncontrollabledisentangledstyle,qi2024deadiffefficientstylizationdiffusion,jun2024disentanglingdisentangledrepresentationsimproved}. These areas highlight the critical innovations that have enabled diffusion models to effectively address the challenges in style transfer.

The primary strength of diffusion models stems from their core mechanism: iteratively estimating the score function, i.e., the gradient of the log-likelihood of the data distribution \cite{song2020generativemodelingestimatinggradients}. This direct estimation of the gradient is a powerful feature, making the noising-denoising process a natural image-to-image framework and, more importantly, allowing for the flexible integration of external, differentiable guidance signals (such as CLIP loss) without requiring model retraining. This principle has given rise to advanced techniques like Score Distillation Sampling (SDS) \cite{alldieck2024scoredistillationsamplinglearned}, which enables diffusion models to act as powerful priors for complex tasks like 3D generation. Furthermore, despite the computational overhead of the iterative process, its well-defined structure is highly amenable to efficient model distillation techniques \cite{zhou2025attentiondistillationunifiedapproach}, which significantly reduce sampling steps and mitigate latency issues.

However, these capabilities come with inherent trade-offs. The potent stochasticity of the denoising process can lead to a critical challenge in style transfer: maintaining a balance between strong stylization and preserving the original image's structure, often resulting in content leakage or geometric distortion. Another limitation arises from the model's holistic generation process, where each update step affects the entire feature map. This makes fine-grained, localized edits difficult, as minor adjustments can cause unintended global changes, often necessitating complex masking techniques or full regeneration. When compared to autoregressive (AR) models, this reveals a complementary relationship: AR models excel at structural fidelity and layout control, whereas diffusion models lead in textural realism and guidance flexibility. This dichotomy suggests a promising future direction in hybrid frameworks that leverage AR models for structural scaffolding and diffusion models for high-fidelity rendering, combining the strengths of both paradigms \cite{hidreami1technicalreport,scaling-diffusion-language-models,tang2024hartefficientvisualgeneration}.

In addition, a comprehensive summary of the methods, techniques, and contributions discussed in this section is provided in Table \ref{tab:diffusion_Summary}, which offers a consolidated reference for researchers and practitioners exploring the application of diffusion models in AI-driven style transfer. Further technical details and in-depth discussions on diffusion models can be found in the Appendix \ref{diffusion_add}.

\definecolor{line-color}{RGB}{175, 131, 148}
 \definecolor{fill-color}{RGB}{114, 200, 222}

\tikzstyle{category}=[
    rectangle,
    draw=line-color,
    rounded corners,
    text opacity=1,
    minimum height=1.5em,
    minimum width=5em,
    inner sep=2pt,
    align=center,
    fill opacity=.5,
]
\tikzstyle{leaf}=[category,minimum height=1.5em,
    fill=fill-color!40, text width=20em,  text=black,align=left,font=\scriptsize,
    inner xsep=2pt,
    inner ysep=1pt,
]

\begin{table*}[h!]
  \centering
  \caption{Key Innovations in Diffusion Models for AI-Driven Style Transfer. This figure presents a structured overview of advancements in diffusion models applied to style transfer, categorized into four key areas: Enhanced Latent Representations for Style Transfer, Optimized Training Processes for Stability and Scalability, Advanced Conditioning Mechanisms for Multi-Modal and Cross-Domain Style Transfer, and Disentanglement and Content Preservation in Diffusion Models. }
\label{tab:diffusion_Summary}
\begin{forest}
  forked edges,
  for tree={
  grow=east,
  reversed=true,
  anchor=base west,
  parent anchor=east,
  child anchor=west,
  base=left,
  font=\small,
  rectangle,
  draw=line-color,
  rounded corners,align=left,
  minimum width=2.5em,
s sep=4pt,
inner xsep=10pt,
inner ysep=2pt,
align=left,
ver/.style={rotate=90, child anchor=north, parent anchor=south, anchor=center},
  },
  where level=1{text width=12.5em,font=\scriptsize}{},
  where level=2{text width=12.4em,font=\scriptsize}{},
  where level=3{text width=1.90em,font=\scriptsize}{},
  [diffusion 
    [Enhanced Latent Representations \\ for Style Transfer
        [Score-Based
            [SBGMLS\cite{NEURIPS2021_5dca4c6b}{,} SBGM-SDE\cite{song2021scorebased}{,} SBGM-CDLD\cite{dockhorn2022scorebased}\\
            GM-EGDD\cite{NEURIPS2019_3001ef25}{,} PFGM\cite{xu2022poissonflowgenerativemodels}{,} 
            VP-DGM-SM\cite{huang2021variationalperspectivediffusionbasedgenerative}\\
            score-based\cite{song2020generativemodelingestimatinggradients}{,} high-dimensional\\
            tasks\cite{dhariwal2021diffusion}{,} Directional Score Matching \cite{mardia2016scorematchingestimatorsdirectional}
            ,leaf,text width=23em]
        ]
        [Hierarchical Latent\\ Representations
            [HRIS-LDM\cite{Rombach_2022_CVPR}{,} Frido-FPD\cite{fan2022fridofeaturepyramiddiffusion}{,} multi-scale latent\cite{nichol2021improveddenoisingdiffusionprobabilistic}\\            ldm\cite{rombach2022highresolutionimagesynthesislatent}{,} DIT\cite{sun2024ecditscalingdiffusiontransformers}{,}Token Transforming\cite{zeng2025tokentransformingunifiedtrainingfree}
            ,leaf,text width=23em]
        ]
        [Latent Embedding Techniques
            [VQDM-TIS\cite{Gu_2022_CVPR}{,} CLIP-guided embeddings \cite{radford2021learningtransferablevisualmodels}\\
            Text-to-image diffusion models \cite{nichol2022glidephotorealisticimagegeneration}{,} Latent Alignment \\
            Models \cite{deng2018latentalignmentvariationalattention}{,} Normalizing Flow Diffusion (NFD) \cite{ho2020denoising}\\
            DP-VAE\cite{ORBi-f53917f2-cec2-4a85-9ed0-8a248cfa595e}{,} StyleTokenizer\cite{li2024styletokenizerdefiningimagestyle}
            ,leaf,text width=23em]
        ]
        [Cross-Domain Style\\Transfer Applications
            [Guided Multi-Modal Diffusion \cite{dhariwal2021diffusion}\\
            sketch-to-painting transformations \cite{ramesh2022hierarchicaltextconditionalimagegeneration}
            ,leaf,text width=23em]
        ]
    ]
    [Optimized Training Processes for\\ Stability and Scalability
        [Advanced Noise Schedules
            [CMCDM\cite{yang2024crossmodal}{,} IDDPM\cite{pmlr-v139-nichol21a}{,} Variational Diffusion\\
            Models\cite{NEURIPS2021_b578f2a5}{,} 
            CCDF-SC\cite{Chung_2022_CVPR}{,} PNM-DMM\cite{liu2022pseudo}
            ,leaf,text width=23em]
        ]
        [Reverse Process Optimizations
            [SDDMDSS\cite{NEURIPS2021_958c5305}{,} ASM-ISIG\cite{jolicoeur-martineau2021adversarial}{,} SGAT-DM\cite{yang2024structureguidedadversarialtrainingdiffusion}\\
            DDIM\cite{Song2020DenoisingDI}{,} gDDIM\cite{zhang2023gddimgeneralizeddenoisingdiffusion}{,} DPM-Solver\cite{lu2022dpmsolverfastodesolver}\\
            FS-DM-EI\cite{zhang2023fastsamplingdiffusionmodels}{,} CFM-Velocity\cite{yang2024consistencyflowmatchingdefining}{,} MLT-SBD-ODE\cite{pmlr-v162-lu22f}\\
            MLT-INDM\cite{kim2022maximum}{,} Rectified Diffusion\cite{wang2024rectifieddiffusionstraightnessneed}
            ,leaf,text width=23em]
        ]
        [Dynamic and \\Adaptive Sampling
            [GGF-SBM\cite{jolicoeurmartineau2021gottafastgeneratingdata}{,} AdaptiveDiffusion \cite{ye2024trainingfreeadaptivediffusionbounded}\\
            Step-Adaptive Training\cite{li2024stepsequalefficientgeneration}{,} AdaDiff \cite{tang2024adadiffacceleratingdiffusionmodels}
            ,leaf,text width=23em]
        ]
        [Hybrid Training Frameworks
            [MDT \cite{gao2024mdtv2maskeddiffusiontransformer}{,} U-ViT \cite{bao2023worthwordsvitbackbone}{,} DiffiT \cite{hatamizadeh2024diffitdiffusionvisiontransformers}{,} DIT \cite{peebles2023scalablediffusionmodelstransformers}\\
            FastDiff 2 \cite{huang2023fastdiff}{,} HLDM-TFIST\cite{masui2024harnessinglatentdiffusionmodel}
            ,leaf,text width=23em]
        ]
        [Distributed Training \\and Scalability
            [ADM-MP\cite{chen2024acceleratingdiffusionmodelsparallel}{,} HEDSDM\cite{ma2024efficientdiffusionmodelscomprehensive}{,} SLDM\cite{ravishankar2024scalingpropertiesdiffusionmodels}
            ,leaf,text width=23em]
        ]
        [Augmented Training for \\Robust Generalization
            [REDAP\cite{lee2023robustevaluationdiffusionbasedadversarial}{,} ARTID\cite{RobustifyT2I}{,} TiNO-Edit\cite{chen2024tino}\\
            Deep Data Consistency\cite{chen2024deep}
            ,leaf,text width=23em]
        ]
        [Enhanced Training Objectives
            [IDIS-CP\cite{yang2024improvingdiffusionbasedimagesynthesis}{,} EDS-DBG\cite{karras2022elucidatingdesignspacediffusionbased}{,} MLT-SBDM\cite{NEURIPS2021_0a9fdbb1}\\
            B-LoRA\cite{frenkel2024implicitstylecontentseparationusing}{,} ZeroShotCL\cite{yang2023zeroshotcontrastivelosstextguided}
            ,leaf,text width=23em]
        ]
    ]
    [Advanced Conditioning\\Mechanisms for Multi-Modal and\\Cross-Domain Style Transfer
        [Attention-Driven\\Conditioning Mechanisms
            [HTCIG-CL\cite{ramesh2022hierarchicaltextconditionalimagegeneration}{,} PTIM-DLU\cite{saharia2022photorealistic}{,} SCEPTER\cite{han2024styleboothimagestyleediting}
            ,leaf,text width=23em]
        ]
        [Domain-Specific Conditioning \\for Cross-Domain Transfer
            [DreamBooth\cite{ruiz2023dreamboothfinetuningtexttoimage}{,} UniTune\cite{valevski2023unitunetextdrivenimageediting}{,} DACDM \cite{zhang2023domainguidedconditionaldiffusionmodel}\\
            Cross-Domain Compositing \cite{hachnochi2023crossdomaincompositingpretraineddiffusion}{,} MagicFace\cite{wang2024magicfacetrainingfreeuniversalstylehuman}\\
            StyleID\cite{zhang2023inversionbasedstyletransferdiffusion}{,} PortraitDiff\cite{zuo2024multiviewconsistentstyletransfer}{,}SCP-Diff\cite{gao2024scp}
            ,leaf,text width=23em]
        ]
        [Hybrid Conditioning \\Architectures
            [CMCDM\cite{yang2023improving}{,} BD-TDE\cite{Avrahami_2022_CVPR}{,} HTCIG-CL\cite{ramesh2022hierarchicaltextconditionalimagegeneration}\\
            GLIDE\cite{nichol2022glidephotorealisticimagegeneration}{,} Imagic\cite{kawar2023imagictextbasedrealimage}{,} MTID-MLLM\cite{yang2024masteringtexttoimagediffusionrecaptioning}\\
            EditWorld\cite{yang2024editworldsimulatingworlddynamics}{,} IterComp\cite{zhang2024itercompiterativecompositionawarefeedback}{,} InstantID\cite{liu2023portraitdiffusiontrainingfreeface}
            ,leaf,text width=23em]
        ]
        [Cross-Modal Style\\Consistency
            [DiffusionCLIP\cite{Kim_2022_CVPR}{,} RealCompo\cite{zhang2024realcompobalancingrealismcompositionality}{,} style transfer\\
            scenarios \cite{saharia2022photorealistictexttoimagediffusionmodels}{,} CreativeSynth\cite{huang2024creativesynthcreativeblendingsynthesis}{,} StyleAligned\cite{zhang2023prospectpromptspectrumattributeaware}
            ,leaf,text width=23em]
        ]
        [Multi-Modal \\Embedding Strategies
            [ALIGN\cite{jia2021scalingvisualvisionlanguagerepresentation}{,} td content generation \cite{ramesh2021zeroshottexttoimagegeneration}{,}
            MUMU\cite{berman2024mumubootstrappingmultimodalimage}
            ,leaf,text width=23em]
        ]
    ]
    [Disentanglement and Content \\Preservation in Diffusion Models
        [Innovative Approaches \\to Disentanglement
            [StyleDiffusion \cite{wang2023stylediffusioncontrollabledisentangledstyle}{,} DEADiff\cite{qi2024deadiffefficientstylizationdiffusion}{,} latent U-Net\cite{ronneberger2015unetconvolutionalnetworksbiomedical}\\
            DRL\cite{jun2024disentanglingdisentangledrepresentationsimproved}
            ,leaf,text width=23em]
        ]
    ]
  ]
\end{forest}
\end{table*}

\subsection{Autoregressive (AR) Image-to-Image Style Transfer}

Autoregressive modeling decomposes the joint likelihood of an image into a sequence of conditional probabilities, enabling pixel-by-pixel generation. This approach was first introduced in the vision domain by PixelRNN \cite{oord2016pixelrecurrentneuralnetworks} and PixelCNN \cite{oord2016conditionalimagegenerationpixelcnn}, where the likelihood is expressed as $p(x)=\prod_{i}p(x_i\mid x_{<i})$. Their $\mathcal{O}(N)$ sampling latency and limited receptive fields hindered high-resolution image synthesis despite being theoretically exact. PixelCNN++ \cite{salimans2017pixelcnnimprovingpixelcnndiscretized} improved upon this with discretized logistic likelihoods and blockwise convolutions, closing the quality gap but still struggling with resolutions beyond $256^2$.

The introduction of Transformer architectures has reignited interest in AR for image generation. MaskGIT \cite{chang2022maskgitmaskedgenerativeimage} tackled the challenge of slow sampling by using random-mask parallel decoding. This approach replaces strict left-to-right pixel generation with multiple rounds of masked-token refinement, enabling near-parallel sampling and controllable diversity. Similarly, Pix2Seq-v2 \cite{chen2021pix2seq} demonstrated that a Transformer-AR backbone can efficiently handle spatial tasks, such as detection, proving its capability for 2-D sequence modeling.

A significant advancement was made with the \textbf{VAR} model \cite{tian2024visualautoregressivemodelingscalable}, which first generates a $32{\times}32$ VQ-GAN latent grid using full-order AR, followed by two super-resolution stages. On ImageNet-512, VAR achieved an impressive FID score of 1.73. Building on this, subsequent methods focused on editing capabilities: EditAR\cite{mu2025editarunifiedconditionalgeneration} refines masked regions with token-wise conditioning, while DAR incorporates direction-aware attention to align brush strokes \cite{xu2025directionaware}. These advancements significantly enhance local fidelity, although sampling latency remains a key bottleneck.

While AR models have shown great promise, diffusion models, operating in pixel or noise space, offer an alternative with a step-wise, differentiable trajectory. This allows external signals (such as CLIP loss, style references, or text prompts) to be injected at every timestep using techniques like classifier-free guidance \cite{ho2022classifierfreediffusionguidance}, ControlNet \cite{zhang2023addingconditionalcontroltexttoimage}, and DreamBooth \cite{ruiz2023dreamboothfinetuningtexttoimage} —without requiring additional training. In contrast, AR decoders sequentially predict discrete latent tokens whose semantics are compressed and indirect. To steer generation in AR models, one must either modify token logits (via logit fusion) or modulate Transformer activations (using style-conditioned embeddings or attention weights), which typically require additional adapters or fine-tuning. These adjustments, however, cannot be applied at arbitrary intermediate stages. To close this gap, future work may need to (i) incorporate flow-matching ODE solvers into AR decoding, (ii) utilize sparse attention or learned token priors to reduce sequence length, and (iii) explore hybrid AR-diffusion frameworks that combine the pixel-level guidance of diffusion with the layout fidelity of AR models. We have summarized some of the recent, representative works in AR modeling in Table \ref{tab:ar_timeline}.

\begin{table}[t!]
    \centering
    \caption{Milestones in AR modelling for vision and their relevance to style transfer}
    \label{tab:ar_timeline}
    \begin{tabular}{p{1cm}p{2cm}p{4.5cm}}
        \toprule
        Year & Key Work & Contribution and Limitation \\ \midrule
        
        2014–16 & PixelRNN \cite{oord2016pixelrecurrentneuralnetworks} / PixelCNN \cite{oord2016conditionalimagegenerationpixelcnn} & First exact pixel-wise likelihood models for AR; however, \(\mathcal{O}(N)\) sampling is too slow for high-resolution and struggles with capturing global semantics. \\
        
        2019 & PixelCNN-HF \cite{salimans2017pixelcnnimprovingpixelcnndiscretized} & Introduced blockwise and sparse attention speed-ups for faster inference; still limited in generating resolutions beyond \(256^2\). \\
        
        2022 & MaskGIT \cite{chang2022maskgitmaskedgenerativeimage} & Random-mask parallel decoding; successfully bridges the quality-speed gap and maintains controllable randomness for efficient sampling. \\
        
        2024 & VAR \cite{tian2024visualautoregressivemodelingscalable} & First AR model to outperform diffusion models at high resolution, achieving state-of-the-art results in generative performance. \\
        
        2024 & Infinity \cite{Infinity2024} & Advanced AR framework that continues to push the boundaries in terms of speed and quality, paving the way for further innovations in high-resolution generation. \\
        
        2025 & EditAR \cite{mu2025editarunifiedconditionalgeneration} / DAR \cite{xu2025directionaware} & Introduces mask-based local editing and direction-aware attention, improving spatial fidelity and enhancing editing flexibility. \\
        \bottomrule
    \end{tabular}
\end{table}

\section{Evaluation Metrics and Key Innovations in Style Transfer}

\subsection{Evaluation Metrics}
Effective evaluation of style-transfer methods necessitates precise metrics capturing essential visual and semantic qualities, including perceptual fidelity, content consistency, output diversity, and realism. Yet, a universally accepted perceptual metric remains elusive due to the inherent complexity of human vision, prompting researchers to develop surrogate computational measures and deep-learning–based approaches. We propose five essential dimensions for evaluating style-transfer performance, aligned closely with recent research trends:
\textbf{Perceived Fidelity (PF, subjective visual authenticity and perceptual similarity)},
\textbf{Content Preservation (CP, retention of semantic and structural information)},
\textbf{Diversity (DIV, stylistic variability and output richness)},
\textbf{Perceptual Quality (PQ, objective approximations of human visual quality)},
and \textbf{Deep Representation Alignment (DRA, neural embedding-based perceptual metrics)}.

Detailed metric definitions, specific computational methods, and comprehensive evaluation protocols for these metrics are systematically outlined in Appendix~\ref{evaluation_add}.

\subsection{Key Techniques and Innovations}
Recent work in style transfer converges on six practical axes: 
\textbf{Artistic Merit (AM, aesthetic fidelity and stylistic authenticity)}, 
\textbf{Visual Quality (VQ, perceptual realism and structural coherence)}, 
\textbf{Computational Efficiency (CE, inference speed and resource footprint)}, 
\textbf{Robustness \& Generalization (R, stable performance across styles, resolutions, and datasets)}, 
\textbf{Multimodal Capability (MC, adaptability to other modalities or cross-domain tasks)}, 
and \textbf{Ethical \& Safety (ES, intellectual-property, misuse, and cultural-sensitivity risks)}. 
We benchmark representative methods from 2023–2025 along these axes (see Table~\ref{tab:contrast}); 
full metric definitions and measurement protocols are provided in Appendix~\ref{app:keytech}.

\begin{table}[t!]
\centering
\caption{The symbols used in the table follow a standardized rating system: ✓ represents ``Excellent'', \checkmark{} indicates ``Generally Good'', and ✗ denotes ``Relatively Low Computational Efficiency''.}
\label{tab:contrast}
\begin{tblr}{
  colspec = {ccccccc},
  row{1} = {bg=gray!30, font=\bfseries},
  row{3} = {bg=gray!10},
  row{5} = {bg=gray!10},
  row{7} = {bg=gray!10},
  row{9} = {bg=gray!10},
  row{11} = {bg=gray!10},
  row{13} = {bg=gray!10},
  row{15} = {bg=gray!10},
  row{17} = {bg=gray!10},
  row{19} = {bg=gray!10},
  row{21} = {bg=gray!10},
  row{23} = {bg=gray!10},
  row{25} = {bg=gray!10},
  row{27} = {bg=gray!10},
}
Method &  \makecell[c]{MC} & \makecell[c]{AM} & \makecell[c]{VQ} &  \makecell[c]{CE} & \makecell[c]{R} & \makecell[c]{Year}\\
InST\cite{zhang2023inversionbasedstyletransferdiffusion} & \usym{2714} & \checkmark &\checkmark & \usym{2717} & \usym{2717} & 2023\\
ZSCL\cite{yang2023zeroshotcontrastivelosstextguided} & \usym{2714} & \checkmark &\checkmark &\usym{2717} & \checkmark & 2023\\
ControlStyle\cite{chen2023controlstyletextdrivenstylizedimage} &\usym{2714} &\checkmark &\usym{2714} &\usym{2717} &\usym{2717} & 2023\\
Artbank\cite{Zhang_Zhang_Xing_Li_Zhao_Sun_Lan_Luan_Huang_Lin_2024} & \usym{2714} & \usym{2717} &\checkmark & \usym{2717} & \usym{2714} & 2023\\
Portrait Diffusion\cite{liu2023portrait} & \usym{2717} & \checkmark & \checkmark & \usym{2717} & \usym{2717} & 2024\\
StyleID\cite{chung2024styleinjectiondiffusiontrainingfree} & \usym{2717} & \checkmark &\usym{2714} & \checkmark & \usym{2714} & 2024\\
diffstyler\cite{li2024diffstyler} & \usym{2717} & \usym{2714} &\usym{2714} & \checkmark & \usym{2714} & 2024\\
StyleTokenizer\cite{li2024styletokenizerdefiningimagestyle} & \usym{2714} & \checkmark & \checkmark & \usym{2717} & \usym{2717} & 2024\\
MagicFace\cite{wang2024magicfacetrainingfreeuniversalstylehuman} & \usym{2714} & \usym{2717} & \checkmark & \usym{2717} & \usym{2717} & 2024\\
SCEPTER\cite{scepter} & \usym{2714} & \checkmark & \usym{2717} & \usym{2717} & \checkmark & 2024\\
B-LoRA\cite{frenkel2024implicitstylecontentseparationusing} & \usym{2717} & \checkmark & \usym{2717} & \usym{2714} & \usym{2717}& 2024\\
CreativeSynth\cite{huang2024creativesynthcreativeblendingsynthesis} & \usym{2714} & \usym{2714} & \usym{2717} & \checkmark & \usym{2717} & 2024\\
Art-Free\cite{ren2024art-free} & \usym{2714} & \usym{2714} & \usym{2714} & \usym{2717} & \usym{2717} & 2024\\
FreeStyle\cite{he2024freestyle} & \usym{2714} & \checkmark &\usym{2714} & \usym{2717} & \checkmark & 2024\\
StyleAligned\cite{hertz2024stylealignedimagegeneration} & \usym{2714} & \usym{2714} & \usym{2714} & \usym{2717} & \usym{2717} & 2024\\
ProSpect\cite{zhang2023prospectpromptspectrumattributeaware} &\usym{2714} & \checkmark & \checkmark & \usym{2717} & \usym{2717} & 2024\\
InstantID\cite{wang2024instantid} & \usym{2717} & \usym{2714} & \usym{2714} & \usym{2714} & \usym{2714} & 2024\\
CSGO \cite{xing2024csgo} &\usym{2714} &\checkmark & \checkmark&\usym{2714} & \checkmark & 2024\\
DiffuseST\cite{hu2024diffusestunleashingcapabilitydiffusion} & \usym{2717} &\usym{2714} &\checkmark & \usym{2717}& \usym{2714}& 2024\\
Z-STAR+\cite{deng2024zstarzeroshotstyletransfer} &\usym{2717} &\usym{2714} &\checkmark &\usym{2717} &\checkmark & 2024\\
ArtAdapter\cite{chen2024artadaptertexttoimagestyletransfer} &\usym{2714} &\usym{2714} &\usym{2714} &\usym{2717} &\checkmark & 2024\\
\end{tblr}
\end{table}

\section{Applications and Practices}

Style transfer today reaches far beyond classic image-to-image tasks, proving useful in five major domains—\textbf{portrait stylisation (PS)}, \textbf{video transformation (VT)}, \textbf{3-D style transfer (3DST)}, \textbf{text style transfer (TST)}, and \textbf{domain adaptation (DA)}.  
A concise definition and representative benchmarks for each domain are given in Appendix~\ref{app:domains}, while a survey of recent studies is summarised in Table~\ref{tab:style_transfer}.

\begin{table*}[h!]
\renewcommand{\arraystretch}{1.2}
\setlength{\tabcolsep}{8pt} 
\centering
\caption{Applications and Practices of Style Transfer.}
\label{tab:style_transfer}
\resizebox{\textwidth}{!}{ 
\begin{tabularx}{\textwidth}{p{3.8cm} X}
\rowcolor{gray!30}
\textbf{Applications} & \textbf{Model} \\
\multirow{4}{*}{Sec \ref{sec:portraits}: Portraits Style Transfer} &
APDrawingGAN\cite{YiLLR19},    
WarpGAN\cite{shi2019warpganautomaticcaricaturegeneration},         
Cartoon-StyleGAN\cite{back2021finetuningstylegan2cartoonface},   
DynaGAN\cite{Kim2022DynaGAN},                  
TargetCLIP\cite{chefer2021targetclip},              
DCT-Net\cite{men2022dct},                           
AiSketcher\cite{gao2020makingrobotsdrawvivid},     
SPatchGAN\cite{Shao_2021_ICCV},          
StyleCariGAN\cite{Jang2021StyleCari},       
CariMe\cite{gu2021carime},                         
BlendGAN\cite{liu2021blendgan},           
MMFS\cite{li2023multimodalfacestylizationgenerative},   
Fix the Noise\cite{Lee_2023_CVPR},             
GODA\cite{Zhang2022GeneralizedOD},               
Mind the Gap\cite{zhu2021mind}, 
CreativeSynth\cite{huang2024creativesynth},
PortraitDiffusion\cite{liu2023portrait},
VCT\cite{cheng2023general},
IP-Adapter\cite{ye2023ip-adapter},
SSR-Encoder\cite{zhang2024ssr},
InstantID\cite{wang2024instantid},
DoesFS\cite{zhou2024deformable},
InstantStyle\cite{wang2024instantstyle},
ZePo\cite{liu2024zepo},
Pair Customization\cite{jones2024customizingtexttoimagemodelssingle}
\\
\midrule

\multirow{3}{*}{Sec \ref{sec:video}: Video Style Transfer} &
ReCoNet\cite{gao2018reconetrealtimecoherentvideo},
Learning Linear Transformations\cite{li2018learning},
VToonify\cite{yang2022Vtoonify}, AdaAttN\cite{liu2021adaattnrevisitattentionmechanism},
Layered Neural Atlases\cite{kasten2021layeredneuralatlasesconsistent},
CCPL\cite{wu2022ccpl}, FateZero\cite{qi2023fatezerofusingattentionszeroshot},
CAP-VSTNet\cite{wen2023capvstnetcontentaffinitypreserved},
Rerender A Video\cite{yang2023rerender},
Style A Video\cite{huang2023style},
Control A Video\cite{chen2023controlavideo},
Adaptive Style Transfer\cite{sanakoyeu2018styleawarecontentlossrealtime},
Hallo\cite{xu2024hallo},
Hallo2\cite{cui2024hallo2},
Hallo3\cite{cui2024hallo3}
\\
\midrule

\multirow{2}{*}{Sec \ref{sec:3d}: 3D Style Transfer} & 
RSMT\cite{10.1145/3588432.3591514}, 
DiffuseStyleGesture\cite{ijcai2023p650},
CAMDM\cite{camdm},
LART\cite{chen2023LART},
Tailor3D\cite{qi2024tailor3dcustomized3dassets},
Style-NeRF2NeRF\cite{fujiwara2024stylenerf2nerf3dstyletransfer},
Dream-in-Style\cite{kompanowski2024dreaminstyletextto3dgenerationusing},
Style3D\cite{song2024style3dattentionguidedmultiviewstyle},
StyleSplat\cite{jain2024stylesplat3dobjectstyle},
Local Motion Phases\cite{mason2022local} \\
\midrule

\multirow{20}{*}{Sec \ref{sec:text}: Text Style Transfer} & 
SETTP\cite{jin2024settpstyleextractiontunable}, 
Sequence to Better Sequence\cite{pmlr-v70-mueller17a},
Language Style Transfer\cite{shen2017styletransfernonparalleltext},
ARAE\cite{2017arXiv170604223J},
Delete, Retrieve, Generate\cite{li2018deleteretrievegeneratesimple},
Sequence to Better Sequence\cite{pmlr-v70-mueller17a},
Controlled Text Generation\cite{hu2017toward},
Cross-Alignment Style Transfer\cite{shen2017style},
Adversarial Autoencoders\cite{zhao2018adversarially},
Zero-Shot Style Transfer\cite{carlson2017zero},
Text Style Transfer Exploration\cite{fu2018style},
Delete, Retrieve, Generate\cite{li2018delete},
SHAPED\cite{zhang2018shaped},
Seq2Seq Sentiment Transfer\cite{singh2018sentiment},
Style Transfer via Back-Translation\cite{prabhumoye2018style},
Unpaired Sentiment Translation\cite{xu2018unpaired},
Offensive Language Transfer\cite{santos2018fighting},
Style Transfer with LM Discriminators\cite{yang2018unsupervised},
Disentangled Representation Learning\cite{john2018disentangled},
Arbitrary Style Transfer\cite{zhao2018language},
Unsupervised MT Style Transfer\cite{zhang2018style},
Sentiment Memory for Style Modification\cite{zhang2018learning},
Multilingual Back-Translation\cite{prabhumoye2018style},
Content Preservation for Style Transfer\cite{tian2018structured},
Controllable Text Formalization\cite{jain2019unsupervised},
Hierarchical Alignment for Text Rewriting\cite{nikolov2018large},
Evaluation Metrics for Style Transfer\cite{pang1810learning},
Attribute Control in Text Generation\cite{logeswaran2018content},
QuaSE\cite{liao2018quase},
Feature-Mover's Distance Generation\cite{chen2018isolating},
Controlled Sentiment Transformation\cite{leeftink2019towards},
Hybrid Formality Style Transfer\cite{xu2019formality},
RL-based Style Transfer\cite{gong2019reinforcement},
Error Correction and Style Transfer\cite{korotkova2019grammatical},
Multi-Attribute Style Transfer\cite{lample2019multiple},
Style Transformer\cite{dai2019style},
Dual RL for Style Transfer\cite{luo2019dual},
Politeness via Side Constraints\cite{sennrich2016controlling},
Formality Control in MT\cite{niu-etal-2017-study},
Controlling Linguistic Style\cite{ficler2017controllinglinguisticstyleaspects},
Writing Style and Fraud\cite{braud2017writingstylepredictivescientific},
Pseudo-Parallel Data for Sequence Editing\cite{Liao2018IncorporatingPD},
Polite Dialogue Generation\cite{niu2018politedialoguegenerationparallel},
Adversarial Text Decomposition\cite{romanov2019adversarialdecompositiontextrepresentation},
Stylish Image Description Generation\cite{chen2018unsupervisedstylishimagedescription},
Lyrics Generation with VAE\cite{vechtomova2018generatinglyricsvariationalautoencoder},
Sentence Editing Prototypes\cite{guu2018generatingsentenceseditingprototypes},
ALTER\cite{xu2019alterauxiliarytextrewriting},
Wasserstein Autoencoders for Stylized Text\cite{ghabussi2019stylizedtextgenerationusing},
Non-Parallel Author-Stylized Rewriting\cite{syed2020adaptinglanguagemodelsnonparallel},
Latent Space Structuring for Response Generation\cite{gao2019structuringlatentspacesstylized},
Label-Conditional Text Generation\cite{Li_Li_Zhang_Li_Zheng_Carin_Gao_2020},
Controlled Headline Generation\cite{jin2020hooksheadlinelearninggenerate},
Contextual Style Relevance\cite{zhou2022exploringcontextualwordlevelstyle},
Formality Data Augmentation\cite{zhang-etal-2020-parallel},
Politeness Transfer\cite{madaan2020politenesstransfertaggenerate},
Profanity Redaction Framework\cite{tran2020friendlyonlinecommunityunsupervised},
Simile Generation via Style Transfer\cite{chakrabarty-etal-2020-generating},
PowerTransformer\cite{ma2020powertransformerunsupervisedcontrollablerevision},
SC2\cite{zhao2024sc2enhancingcontentpreservation},
BERT and Linguistic Styles\cite{hayati2021doesbertlearnhumans}
\\
\midrule

\multirow{7}{*}{Sec \ref{sec:domain}: Domain Adaptation} &
One-Shot Domain Adaptation for Face Generation\cite{yang2020oneshotdomainadaptationface},
Meta Face Recognition in Unseen Domains\cite{guo2020learningmetafacerecognition},
Cross-Domain Document Detection\cite{li2020crossdomaindocumentobjectdetection},
StereoGAN\cite{liu2020stereoganbridgingsynthetictorealdomain},
FISC\cite{nguyen2024fiscfederateddomaingeneralization},
Domain Adaptation for Image Dehazing\cite{9156566},
Probability Weighted Features for Retrieval\cite{huang2020probabilityweightedcompactfeature},
Disparity-Aware Domain Adaptation\cite{9156713},
PointDAN\cite{qin2019pointdanmultiscale3ddomain},
Head Pose Estimation with Adversarial Adaptation\cite{9009467},
Cross-Domain Animal Pose Estimation\cite{cao2019crossdomainadaptationanimalpose},
GA-DAN\cite{zhan2019gadangeometryawaredomainadaptation},
Accelerating Unsupervised Domain Adaptation\cite{yu2019acceleratingdeepunsuperviseddomain},
Cycle-consistent Adversarial Transfer\cite{li2019cycleconsistentconditionaladversarialtransfer},
GCAN\cite{8953825},
CSIT-DGMI\cite{spanos2024complexstyleimagetransformations},
Image Translation for Domain Adaptation\cite{murez2017imageimagetranslationdomain},
Conditional GAN for Domain Adaptation\cite{8578243},
Domain Conditioned Adaptation Network\cite{li2020domainconditionedadaptationnetwork},
Deep Transfer with Joint Adaptation\cite{long2017deeptransferlearningjoint},
SDNIA-YOLO\cite{ding2024sdniayolorobustobjectdetection},
Augmenting Synthetic Images for Sim2Real\cite{pashevich2019learningaugmentsyntheticimages}
\\
\bottomrule
\end{tabularx}
}
\end{table*}

\section{Datasets and Evaluation Criteria}

\subsection{Construction and Analysis of Datasets}

Datasets are essential for developing and benchmarking style transfer models. In the anime domain, the Danbooru series\cite{danbooru2018,danbooru2020,danbooru2021} provides extensive annotated image collections for style learning. Video style transfer leverages datasets such as UADFV\cite{yang2018exposingdeepfakesusing}, EBV\cite{li2018ictuoculiexposingai}, and Deepfake-TIMIT\cite{zi2024wilddeepfakechallengingrealworlddataset} to support tasks such as deepfake detection and expression swapping. For text style transfer, datasets such as YAFC Corpus\cite{rao2018dearsirmadami} and StylePTB\cite{lyu2021styleptbcompositionalbenchmarkfinegrained} enable sentiment and politeness transfer, while ParaDetox ensures ethical compliance. In 3D style transfer, datasets such as 100STYLE\cite{mason2022local} and Bandai-Namco-Research\cite{kobayashi2023motion} facilitate artistic transformations of 3D models. 

The diversity of these datasets has driven significant progress in style transfer. A detailed discussion is provided in the Appendix \ref{Datasets}, and Table \ref{tab:Datasets} summarizes datasets across Anime, Video, Text, and 3D style transfer.

\begin{table*}[h!]
\renewcommand{\arraystretch}{1.2}
\setlength{\tabcolsep}{8pt}
\centering
\caption{Overview of datasets used for Style Transfer.}
\label{tab:Datasets}
\resizebox{\textwidth}{!}{
\begin{tabular}{l l l l}
\rowcolor{gray!30}
\textbf{Year} & \textbf{Dataset} & \textbf{Size} & \textbf{Description} \\
\midrule
\rowcolor{gray!20} \multicolumn{4}{l}{\textbf{Anime Style Transfer}} \\
\rowcolor{white} 2017 & Danbooru2017 & 1.9TB (2.94M images, 77.5M tags) & Anime \\
\rowcolor{gray!10} 2018 & Danbooru2018 \cite{danbooru2018} & 2.5TB (3.33M images, 92.7M tags) & Anime \\
\rowcolor{white} 2019 & Danbooru2019 & 3TB (3.69M images, 108M tags) & Anime \\
\rowcolor{gray!10} 2020 & Danbooru2020 \cite{danbooru2020} & 3.4TB (4.23M images, 130M tags) & Anime \\
\rowcolor{white} 2021 & Danbooru2021 \cite{danbooru2021} & 4.9M images, 162M tags & Anime \\
\rowcolor{gray!10} 2022 & Danbooru2022 & 4M+ images & Anime \\
\rowcolor{white} 2023 & Danbooru2023 & 8TB & Anime \\
\rowcolor{gray!10} 2018 & Chinese Style Transfer \cite{ChineseStyle} & 1K content, 100 style images & Chinese Painting \\
\rowcolor{white} 2018 & Stylized ImageNet \cite{geirhos2018} & 134GB & Style Transfer \\
\rowcolor{gray!10} 2018 & WikiArt \cite{mancini2018addingnewtaskssingle} & 42,129 images & Style, Artist, and Genre Classification \\
\rowcolor{white} 2019 & FFHQ \cite{karras2019style} & 70K images & Human Faces \\
\rowcolor{gray!10} 2019 & Dark Zurich Dataset \cite{SDV19} & 8,779 images & Night Day Captures \\
\rowcolor{white} 2020 & Comic Faces & 20K images & Paired Comic Faces \\
\rowcolor{gray!10} 2020 & iFakeFaceDB \cite{Neves_2020} & 87K synthetic faces & Face Dataset \\
\rowcolor{white} 2020 & Ukiyo-e Faces \cite{pinkney2020ukiyoe} & 5,209 images & Aligned Ukiyo-e Faces \\
\rowcolor{gray!10} 2020 & DFFD \cite{dang2020detectiondigitalfacemanipulation} & 299K images & Digital Face Manipulation \\
\rowcolor{white} 2020 & Cartoon Faces \cite{pinkney2020resolutiondependentganinterpolation} & 35.7M images & Cartoon (Disney) Faces \\
\rowcolor{gray!10} 2020 & MetFaces \cite{karras2020traininggenerativeadversarialnetworks} & 1,336 images & Artistic Face Images \\
\rowcolor{white} 2021 & AAHQ \cite{liu2021blendganimplicitlyganblending} & 25K images & Artistic Faces \\
\rowcolor{gray!10} 2022 & DiffusionDB \cite{wangDiffusionDBLargescalePrompt2022} & 14M images & T2I Prompt Dataset \\
\rowcolor{white} 2022 & StyleGAN Human \cite{fu2022styleganhuman} & 40K+ images & Human Generation \\

\rowcolor{gray!10} 2023 & 4SKST \cite{Seo2023semi} & 25 color, 100 sketches & Sketch Style \\
\rowcolor{white} 2023 & JourneyDB \cite{sun2023journeydbbenchmarkgenerativeimage} & 4.43M images & Generative Image Benchmark \\
\rowcolor{gray!10} 2023 & DeepFakeFace \cite{song2023robustnessgeneralizabilitydeepfakedetection} & 12K images & Artificial Celebrity Faces \\
\rowcolor{white} 2024 & DiffusionFace \cite{chen2024diffusionface} & 600K images & Face Forgery \\
\rowcolor{gray!10} 2024 & Trailer Faces HQ \cite{pinkney2023tfhq} & 187K face images & Facial Expressions \\
\rowcolor{white} 2024 & Diffusion Deepfake \cite{bhattacharyya2024diffusion} & 1.2M images & Deepfake Image Generation \\
\rowcolor{gray!10} 2024 & DF40 \cite{yan2024df40} & 40 synthesis techniques & Comprehensive Deepfake Dataset \\
\midrule
\rowcolor{gray!20} \multicolumn{4}{l}{\textbf{Video Style Transfer}} \\
\rowcolor{white} 2018 & UADFV \cite{yang2018exposingdeepfakesusing} & 100 videos & Video Style Transfer \\

\rowcolor{gray!10} 2018 & Deepfake-TIMIT \cite{korshunov2018deepfakesnewthreatface} & 960 videos & Face Recognition \\
\rowcolor{white} 2021 & FFIW-10K\cite{Zhou_2021_CVPR} &10K videos &face forgery detection techniques  \\

\rowcolor{gray!10} 2020 & DFDC \cite{dolhansky2020deepfakedetectionchallengedfdc} & 100K clips & DeepFake Detection \\
\rowcolor{white} 2020 & Celeb-DF (v1)/(v2) \cite{Celeb_DF_cvpr20} & 408 videos & DeepFake \\
\rowcolor{gray!10} 2020 & Deeper Forensic\cite{jiang2020deeperforensics1} & 60,000 videos & Real-World Face Forgery Detection  \\
\rowcolor{white} 2020 & FaceShifter\cite{li2020advancing} & All 1000 original videos & swapped face  \\
\rowcolor{gray!10} 2024 & Wild Deepfake \cite{zi2024wilddeepfakechallengingrealworlddataset} & 7,314 face sequences & Deepfake Detection \\

\midrule
\rowcolor{gray!20} \multicolumn{4}{l}{\textbf{3D Style Transfer}} \\
\rowcolor{white} 2022 & 100STYLE \cite{mason2022local} & 4M frames & 3D Motion Capture \\
\rowcolor{gray!10} 2023 & Bandai-Namco-Research \cite{kobayashi2023motion} & 36.7K frames & 3D Style Transfer \\
\midrule
\rowcolor{gray!20} \multicolumn{4}{l}{\textbf{Text Style Transfer}} \\
\rowcolor{white} 2020 & Touchdown \cite{chen2020touchdownnaturallanguagenavigation} & 9,326 examples & Natural Language Navigation \\
\rowcolor{gray!10} 2020 & Yelp & 6.99M comments & Sentiment and Style Transfer \\
\rowcolor{white} 2018 & YAFC Corpus \cite{rao2018dearsirmadami} & Largest corpus for stylistic transfer & Formality Style Transfer \\
\rowcolor{gray!10} 2020 & ParaDetox \cite{logacheva-etal-2022-paradetox} & 10K sentences & Detoxification with Parallel Data \\
\rowcolor{white} 2022 & TextBox \cite{tang-etal-2022-textbox} & 47 pre-trained LMs & Text Style Transfer \\

\hline

\end{tabular}
}
\end{table*}

\subsection{Data Augmentation}

Data augmentation plays a vital role in enhancing the performance and robustness of style transfer systems. Early methods, such as STaDA \cite{Zheng_2019} and CycleGAN-based emotion augmentation \cite{Bao2019CycleGANBasedES}, demonstrated the use of style transfer itself for data enhancement. Over time, more advanced techniques such as AugMix \cite{hendrycks2020augmixsimpledataprocessing} and Albumentations \cite{Buslaev_2020} have further improved model robustness and expanded augmentation strategies.

Recent trends emphasize multi-modal augmentation, exemplified by AugLy \cite{papakipos2022auglydataaugmentationsrobustness}, which supports diverse and comprehensive training data. These advancements have significantly increased the versatility and performance of style transfer models. Detailed methods are provided in the Appendix \ref{Data Augmentation}. Table \ref{tab:data_augmentation_methods} summarizes key data augmentation techniques.

\begin{table*}[h!]
\caption{Overview of Data Augmentation Methods for AI Style Transfer.}
\label{tab:data_augmentation_methods}
\begin{tblr}{
  colspec = {clll},
  row{1} = {bg=gray!30, font=\bfseries},
  row{7-9} = {bg=gray!10},
  hline{2,7,10,11} = {1-4}{gray!60},
}

Year & Method & Publication & Innovation \\
~ & STaDA \cite{Zheng_2019} & VISIGRAPP & Utilizes style transfer to augment training data \\
~& CycleGAN-Emotion Augmentation \cite{Bao2019CycleGANBasedES} & VISIGRAPP & Employs CycleGAN improve speech emotion recognition \\
2019& Cross-Domain NST Augmentation \cite{xu2019crossdomainimageclassificationneuralstyle} & arXiv & Applies neural style transfer for data augmentation \\
    
~& InstaBoost \cite{fang2019instaboostboostinginstancesegmentation} & arXiv & A copy-pasting method enhancing instance segmentation models \\
~& CutMix \cite{yun2019cutmix} & ICCV & A strategy blending patches and labels for robust classifiers \\
~ & Parallel Formality Augmentation \cite{zhang-etal-2020-parallel} & ACL & Novel data augmentation methods for formality style transfer \\
2020& AugMix \cite{hendrycks2020augmixsimpledataprocessing} & arXiv & A technique improving model robustness and uncertainty estimation \\
~ & Albumentations \cite{Buslaev_2020} & Information & A flexible image augmentation library improving model training \\
2022 & AugLy \cite{papakipos2022auglydataaugmentationsrobustness} & arXiv & A data augmentation library enhancing model robustness across modalities \\
\end{tblr}
\centering
\end{table*}

\section{Discussion}

This review has explored the evolution of AI-driven style transfer, from early Neural Style Transfer (NST) methods to state-of-the-art diffusion models. Our analysis highlights significant progress across key dimensions: Effectiveness, Adaptability, Practicality, Aesthetics, Control, and Ethics. These aspects define the foundation for systematic evaluation and drive innovations in style transfer research. Systematic evaluation metrics reveal both advancements and persistent challenges, categorized into Perceived Distance, Perceptual Metrics, Content Retention, Diversity Metrics, Deep Learning-Based Metrics, and Trade-offs \& Balance. The field has expanded beyond traditional image-to-image transformations to incorporate text, 3D graphics, and video, driven by the increasing adaptability of generative architectures. While GAN- and diffusion-based models have greatly improved generation quality and style control, challenges remain in computational efficiency, robustness to diverse inputs, and fine-grained style manipulation techniques. Integrating multiple generative approaches with advances in multimodal learning has demonstrated potential in addressing these issues. Additionally, as style transfer technologies gain traction in real-world applications, considerations surrounding ethical implications and responsible AI development will become increasingly critical.

\subsection{Technical Analysis and Methodological Insights}

To better understand the evolution of style transfer, it is crucial to analyze how different techniques have built upon one another:

\begin{enumerate}[leftmargin=*]
\item Neural Style Transfer and Adaptive Instance Normalization: NST pioneered the use of CNNs for content and style disentanglement, providing a foundational framework for style transfer. Adaptive Instance Normalization (AdaIN) improved this process by introducing fine-grained control, enabling real-time style manipulation. However, balancing content fidelity and semantic consistency remains challenging, particularly in complex, high-frequency images.

\item Variational Autoencoders and Generative Adversarial Networks: VAEs and GANs advanced generative image modeling, offering distinct advantages. VAEs facilitate structured latent space learning, ensuring content consistency, while GANs achieve superior realism but are prone to mode collapse, restricting output diversity. Despite these advancements, both models struggle to balance computational efficiency, style diversity, and training stability.

\item Diffusion Models and Their Advantages Over GANs:
Diffusion models have redefined image generation with iterative denoising, producing high sample quality and improved training stability. Unlike GANs, diffusion models excel in consistency and fine-detail reconstruction, but they remain constrained by computational demands, especially for high-resolution images. While diffusion models mitigate mode collapse and improve sample diversity, their reliance on iterative denoising remains a fundamental limitation for real-time applications. By contrast, GANs, despite their instability, offer significantly lower inference latency, making them more practical for certain deployment scenarios.

\item Multimodal Learning and Self-Supervised Techniques: Beyond unimodal generative models, the integration of multimodal learning, particularly text-to-image and cross-modal training, is reshaping how models perceive and manipulate style information across diverse domains. By leveraging both visual and textual features, these approaches enhance semantic alignment, improve content-style coherence, and enable cross-domain adaptation.

\end{enumerate}

\section{Conclusion}

The field of style transfer and generative modeling is undergoing a rapid transformation, driven by advances in sample quality, generation diversity, and training stability. However, several scientific and technical challenges remain, shaping the trajectory of future research.

Future research should prioritize three critical directions to address existing challenges and unlock new possibilities. First, reducing the overhead of diffusion models and GANs remains a critical challenge, especially for high-resolution generation. Innovations in model acceleration and efficient training strategies will be pivotal for real-world deployment. Second, integrating visual and textual data enhances semantic consistency, allowing for more precise and controllable style transformations. Future research should explore better alignment mechanisms across modalities to improve contextual coherence. Third, leveraging the strengths of diverse generative architectures, including hybrid approaches that integrate VAEs, GANs, and diffusion models, will be essential to build robust and adaptable frameworks for style transfer.

Beyond artistic and entertainment applications, style transfer is poised to make significant contributions to scientific visualization, data augmentation, and human-computer interaction. These advancements will facilitate real-time content adaptation, enhance interpretability in AI-driven design, and expand cross-disciplinary applications in areas such as virtual reality, medical imaging, and automated content generation. As these challenges are addressed, style transfer and generative modeling will continue to redefine artistic expression, computational creativity, and industrial applications, unlocking novel possibilities across diverse industries.

\small
\bibliographystyle{IEEEtran}
\bibliography{ref}

\clearpage

\section{Development and Application of Generative Models in Style Transfer}
\subsection{VAE-Based Style Transfer}
\label{vae_add}

Standard Variational Autoencoders (VAE) assume a simple Gaussian prior distribution for the latent space, which, while effective for many tasks, imposes significant limitations when dealing with complex style features. Style transfer tasks require not only preserving the content structure of an image but also effectively transferring various styles, which necessitates a latent space with high flexibility and expressive power. However, the Gaussian prior in standard VAEs often fails to capture the diversity and complexity of styles. As a result, much research has focused on improving the latent space structure to enhance VAE's performance in style transfer, particularly in latent space modeling and style control.

\subsubsection{Improved Latent Space Structure}
\label{sec:subsection_3.1.1}

\textbf{Gaussian Mixture Models and Hierarchical Latent Spaces.} Gaussian Mixture VAE (GMVAE) is one of the improvements to the traditional VAE latent space assumption. Dilokthanakul et al. (2016) first proposed GMVAE\cite{dilokthanakul2016deep}, in which the latent space is represented as a mixture of multiple Gaussian components, enabling the model to smoothly transition content between different styles. This model solves the problem of style diversity in style transfer. Further enhancements to GMVAE include Choi et al. Hierarchical GMVAE\cite{choi2018hierarchical}, which introduces multiple layers in the latent space to capture different levels of style information, enhancing the sense of hierarchy in the style transfer process, especially in image-to-image style transfer, such as the transition from oil painting to modern art. Zhang et al. proposed Conditional GMVAE\cite{zhang2020conditional} based on this, which introduces additional conditional variables (such as style tags or content descriptions) in the latent space to make style transfer more controllable and provide finer style adjustment capabilities. Furthermore, Huang et al. combined Conditional GMVAE with the Self-Attention mechanism\cite{huang2021conditional}, leading to significant advances in artistic style transfer (e.g., from Impressionism to modern art styles). This method further refined the levels and details in style transfer, enhancing both the model's expressiveness and the transfer results.

\textbf{Normalizing Flow VAE: Enhancing Latent Space Representation Power.} Another important innovation is the introduction of Normalizing Flows (NF)\cite{rezende2015variational} to enhance the expressive power of the latent space in VAE, which uses a series of reversible transformations to increase the flexibility of the latent space, enabling the model to adapt to more complex data distributions. Normalizing Flow VAE effectively distinguishes between content and style in the style transfer task, thereby achieving more precise style control. Kingma et al. extended this work by introducing deeper reversible transformation networks such as RealNVP\cite{dinh2017real} and Glow\cite{kingma2018glow} to further improve the generation ability of VAE, especially when dealing with complex styles such as abstract art or impressionism. Yang et al. further verified the application of Normalizing Flow VAE in style transfer\cite{yang2020hierarchical}, where hierarchical flow transformation enables the model to control content and style representations in different latent space dimensions, making the style transfer process more flexible. Despite some progress, computational complexity remains a challenge. To solve this problem, Zhao et al. proposed Fast Normalizing Flow\cite{zhao2021fast}, which reduces computational overhead by optimizing the network architecture, making this method more effective for large-scale image generation tasks, especially for real-time style transfer applications such as video style transfer and live art creation.

\subsubsection{Disentanglement Learning and Efficient Implementations in Style Transfer}
\label{sec:subsection_3.1.2}

\textbf{Disentanglement Learning} is a key technique in style transfer, addressing how to effectively separate content and style so that style transfer can be precisely controlled without altering the content structure. Beta-VAE\cite{higgins2017beta} adjusted the KL divergence term in the ELBO to strengthen the decoupling of content and style, achieving significant results in disentangling. Choi et al. proposed multi-scale disentanglement Beta-VAE\cite{choi2018multi}, which decouples content and style at different scales, making style and detail adjustment more precise. They then further combined conditional generative networks with disentanglement learning to propose a new disentanglement method\cite{choi2020disentangling} that can not only control higher-level artistic styles but also accurately adjust finer details (such as brushstrokes and textures). However, disentanglement learning often comes at the cost of reconstruction quality. To address this issue, Chen et al. proposed an Adaptive Disentanglement Method\cite{chen2020adaptive}, which dynamically adjusts the KL divergence weight and reconstruction error, optimizing the separation of content and style while maintaining high-quality reconstruction.

\textbf{Efficient Implementation with Discrete Latent Space for Style Transfer}
VQ-VAE (Vector Quantized VAE)\cite{oord2017neural} solves the "collapse" problem of traditional VAE by discretizing the latent space, making the generated images clearer and more suitable for the expressiveness of style transfer tasks. The model discretizes the latent space, so that the style features can be expressed more clearly. Razavi et al. demonstrated the advantages of VQ-VAE in artistic style transfer\cite{razavi2019generative}, particularly in handling complex textures and the transformation of artistic styles, with VQ-VAE effectively preserving the details of the target style. Liu et al. introduced Conditional VQ-VAE\cite{liu2020conditional}, combining style tags or other conditional inputs with VQ-VAE to enhance the controllability of style transfer, particularly in multi-style generation tasks, enabling more flexible style transformations.

\subsubsection{Adversarial Generative Models}
\label{sec:subsection_3.1.3}

VAE-GAN (Variational Autoencoder Generative Adversarial Network)\cite{larsen2015autoencoding} is an innovative model that combines VAE with a Generative Adversarial Network (GAN). VAE learns the data distribution in the latent space, while GAN uses adversarial training to optimize the generator to generate visually attractive and detailed images. By introducing the discriminator as a perceptual loss, the model improves the quality of generated images, especially in terms of texture and details.

To further improve the style control in VAE-GAN, Yang et al. proposed VAE-GAN+\cite{yang2018vae}, which integrates a Conditional Generative Network. This extension allows users to precisely control the generated images' style and details using additional inputs like style tags or content descriptions.

In addition to the traditional VAE-GAN, another improvement is Wasserstein VAE-GAN (WAE-GAN)\cite{tolstikhin2017wasserstein}, which replaces the KL divergence in traditional VAEs with Wasserstein distance, addressing the "collapse" problem often encountered during training. In style transfer, WAE-GAN generates smoother latent space interpolations and more consistent style transfers, significantly improving image diversity and consistency.

To enhance the style control further, Wang et al. introduced Multi-Scale VAE-GAN\cite{wang2020paintgan}, which uses multi-scale feature learning to capture style information at different levels. By optimizing multi-scale latent representations, the generator can perform style transfer on multiple scales simultaneously, providing finer control. Conditional VAE-GAN\cite{ghosal2020conditional} is an extension of VAE-GAN, where conditional information (e.g., style tags or target content) is integrated to enhance the control of the style transfer process.

\subsection{GAN-Based Style Transfer}
\label{gan_add}
Generative Adversarial Networks (GANs) have become one of the most powerful tools for style transfer, thanks to their ability to learn and generate realistic images through an adversarial framework. Despite their effectiveness, GANs still face several challenges in style transfer tasks, including maintaining content consistency, handling complex styles, and improving training stability.

\subsubsection{Latent Space Modeling for Style Control}
\label{sec:subsection_3.2.1}

The core challenge of style transfer lies in effectively manipulating the latent space to separate content and style while ensuring high-quality results. Recent innovations in latent space modeling have enhanced the flexibility of GANs and improved control over style transfer, allowing for better content preservation and finer manipulation of style.

\begin{figure*}[h]    
    \centering
    \includegraphics[width=1\linewidth]{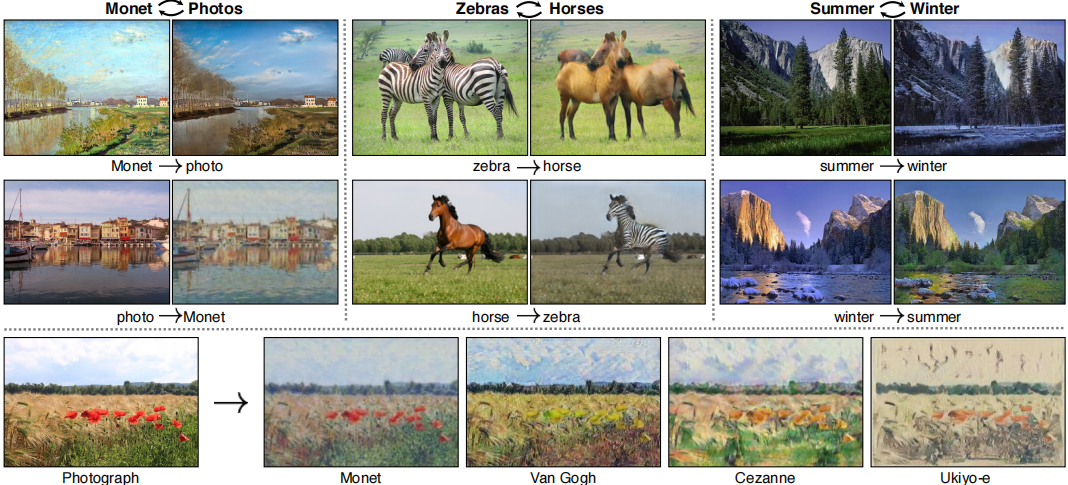} 
    \caption{Given any two unordered sets of images X and Y, CycleGAN\cite{zhu2017unpaired} learns to automatically "convert" images from one to the other and vice versa.}
    \label{fig:cyclegan}
    \vspace{-0.4cm}
\end{figure*}

CycleGAN \cite{zhu2017unpaired}introduces cycle-consistency loss to ensure that images can be converted between different domains (e.g., from photos to paintings) and then returned to their original form without losing key content features. A key improvement is its ability to handle unpaired datasets, which makes it especially useful in artistic style transfer tasks where paired content and style data are not available. Figure \ref{fig:cyclegan} shows the performance of CycleGAN in this regard.

StyleGAN adopts a style-based generator architecture\cite{karras2019style}, where latent vectors are mapped to style vectors using a mapping network and then injected into different layers of the generator. This allows fine-grained control over the style at multiple levels, ranging from coarse content features (such as pose and background) to fine-grained details (such as texture and lighting), providing independent control over high-level content and low-level style features. Users can independently adjust elements like brushstrokes and shading while preserving the underlying content intact. StyleGAN2\cite{karras2019stylegan2} further improves upon this by reducing artifacts, enhancing image quality, and ensuring better detail preservation, making it more effective for style transfer. Similarly, StyleGAN3\cite{karras2021aliasfreegenerativeadversarialnetworks} builds on this by interpreting all signals in the network as continuous to ensure that unwanted information does not leak into the layered synthesis process. StarGAN\cite{choi2018stargan} introduces a unified style transfer framework that supports style conversion across multiple domains, making style transfer more flexible and versatile.

GAT \cite{yuan2024graphattentiontransformernetwork} introduces a graph attention mechanism, further enhancing style control in the latent space and allowing for more detailed adjustments based on both local and global image features. To improve style transfer quality, DualGAN \cite{yi2018dualganunsupervisedduallearning} proposes a bidirectional GAN structure, effectively improving image generation on unpaired datasets. AttentionGAN \cite{tang2021attentionganunpairedimagetoimagetranslation} leverages attention mechanisms to precisely control style features, thereby enhancing image quality during the style transfer process, especially in complex style tasks. PGGAN \cite{karras2018progressivegrowinggansimproved} trains GANs by progressively increasing the resolution, significantly improving the generation of high-resolution images, which is crucial for detail preservation in style transfer. Finally, FUNIT \cite{liu2019fewshotunsupervisedimagetoimagetranslation} proposes a mapping-based style transfer method that enables more precise style control, facilitating cross-domain style transfer and further advancing the diversity and precision of style transfer technologies.

\subsubsection{Training Stability and Enhanced Image Quality}
\label{sec:subsection_3.2.2}

GANs often face instability issues during training, which can result in poor image quality or training failures. This challenge is particularly pronounced in style transfer tasks, where generated images may suffer from mode collapse, gradient vanishing, or exploding gradients. To address these issues, numerous studies have focused on improving training stability and enhancing the quality of generated images.

Wasserstein GAN (WGAN) \cite{arjovsky2017wasserstein} introduced the Wasserstein distance as a replacement for the traditional Jensen-Shannon divergence, significantly improving the stability of GAN training. This innovation provided more consistent training results for style transfer tasks, reducing artifacts in generated images and improving content preservation. To further enhance stability, Spectral Normalization \cite{miyato2018spectralnormalizationgenerativeadversarial} normalized the weights of the discriminator, preventing issues such as gradient explosion and vanishing. By stabilizing the training of both the generator and discriminator, spectral normalization ensured high-quality outputs in style transfer tasks.

Relativistic GAN \cite{jolicoeurmartineau2018relativisticdiscriminatorkeyelement} proposed a relative discriminator method for adversarial training, aimed at improving stability and reducing mode collapse. Instead of directly determining whether samples are real or fake, RaGAN computes relative adversarial information, better balancing the generator and discriminator. This approach is particularly effective in style transfer tasks, ensuring that stylistic features are accurately transferred.

Furthermore, the Laplacian Pyramid GAN \cite{denton2015deepgenerativeimagemodels} introduced a pyramid structure to progressively generate high-resolution images, avoiding instability issues in high-resolution training. This method effectively enhanced image details and quality, which are crucial in style transfer tasks that demand high resolution and detail preservation. Similarly, Progressive GAN \cite{karras2018progressivegrowinggansimproved} proposed a progressive training strategy, generating images layer-by-layer from low to high resolution. This approach prevented instability during high-resolution training and significantly improved image quality.

VAE-GAN \cite{larsen2015autoencoding}  combined the strengths of Variational Autoencoders and GANs, introducing a novel adversarial framework. High-Fidelity GAN \cite{wang2024highfidelityganinversionimage} introduced a new loss function to balance style and content transfer, ensuring high-quality and accurate image transformations. Gradient Penalty GAN \cite{gulrajani2017improvedtrainingwassersteingans} addressed training stability by introducing a gradient penalty term, ensuring gradient updates were neither too large nor too small. This prevented mode collapse and made the model suitable for high-quality image generation tasks. BigGAN \cite{brock2019largescalegantraining} further advanced GAN-based image generation through large-scale training and optimization, significantly enhancing image quality. In style transfer tasks, BigGAN excelled in generating detailed and stylistically accurate images.

\subsubsection{Disentanglement Learning for Independent Content and Style Control}
\label{sec:subsection_3.2.3}

Disentanglement learning plays a critical role in ensuring that content and style factors in the latent space can be independently controlled, which is a key requirement for precise style transfer.

Beta-VAE \cite{higgins2017beta} introduced a mechanism to encourage stronger separation between content and style by scaling the KL divergence term in the ELBO objective function. While initially developed for VAEs, the principles of Beta-VAE have influenced GAN-based models as well. In style transfer tasks, the ability to separate content from style ensures that each element can be independently modified, maintaining content integrity while adjusting styles flexibly.

InfoGAN \cite{chen2016infoganinterpretablerepresentationlearning} further improved disentanglement by maximizing the mutual information of variables in the latent space. Conditional GAN\cite{mirza2014conditionalgenerativeadversarialnets} has been a cornerstone technology, advancing the application of disentanglement learning. By introducing conditional labels (e.g., style labels or content descriptions) during the generation process, cGAN enables precise control over style and content, allowing each characteristic to be independently adjusted. AC-GAN \cite{odena2017conditionalimagesynthesisauxiliary} further enhances this concept by optimizing the relationship between the conditional generator and the discriminator, improving the independence of style and content features.

In addition, DiscoGAN \cite{kim2017learningdiscovercrossdomainrelations} introduced a cross-domain style transfer approach by sharing features in the latent space. This allows content and style features to be independently manipulated while retaining style properties across different domains. This approach is particularly useful for handling cross-domain style transfer tasks, offering greater flexibility in image generation.

Adversarial Autoencoders \cite{makhzani2016adversarialautoencoders} proposed a new disentanglement method by combining GANs with autoencoders. This approach enhances the independence of content and style in style transfer tasks, enabling individual adjustment of each element and optimizing the accuracy of style transfer. Deformation GAN \cite{zhang2020deformgananunsupervisedlearningmodel} introduced a deformation network to establish a finer disentanglement mechanism between style and content features.

MUNIT \cite{huang2018multimodal} proposed a method to control content and style features independently by introducing multiple latent variables. This model excels in cross-domain style transfer, enabling high-quality image transformation by preserving content features while flexibly adjusting styles. Similarly, FUNIT \cite{liu2019fewshotunsupervisedimagetoimagetranslation} introduced a mapping-based style transfer method, facilitating a finer disentanglement of content and style features for more precise control.

\subsection{Diffusion-Based Style Transfer}
\label{diffusion_add}
Diffusion models have emerged as a transformative framework in generative modeling, demonstrating remarkable capabilities in image synthesis and style transfer. 

\subsubsection{Enhanced Latent Representations for Style Transfer}
\label{sec:subsection_3.3.1}

\textbf{Score-Based Diffusion Models:}
The score-based methods introduced by Song et al. \cite{song2020generativemodelingestimatinggradients} established foundational techniques for noise reduction and manifold alignment using Score Matching with Langevin Dynamics and Denoising Score Matching. These approaches iteratively map noisy inputs to refined outputs, producing superior results in high-dimensional tasks like style transfer. Dhariwal and Nichol \cite{dhariwal2021diffusion} extended these ideas in guided diffusion models, incorporating style-specific constraints to balance stylistic transformation and content fidelity. Further developments, such as Directional Score Matching \cite{mardia2016scorematchingestimatorsdirectional}, allow style-specific diffusion by orienting score gradients toward style manifolds. 

\begin{figure}[h!]
    \centering
    \includegraphics[width=0.48\textwidth]{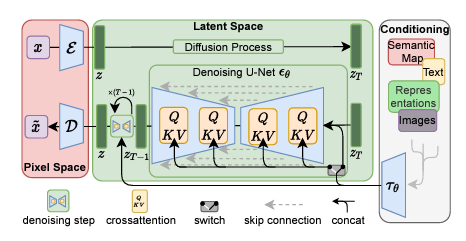}
    \caption{The condition LDM \cite{rombach2022highresolutionimagesynthesislatent} either via concatenation or by a more general cross-attention mechanism.}
    \label{fig:ldm}
\end{figure}

\textbf{Hierarchical Latent Representations:} Hierarchical approaches have introduced multi-scale latent representations, significantly improving adaptability in style transfer. Nichol et al. \cite{nichol2021improveddenoisingdiffusionprobabilistic} and Rombach et al. \cite{rombach2022highresolutionimagesynthesislatent} leveraged these representations in Latent Diffusion Models (LDMs), As shown in Figure \ref{fig:ldm}, segmenting latents into coarse-grained and fine-grained layers. Coarse layers capture global stylistic elements like color schemes, while fine-grained layers preserve textures and intricate details.

\textbf{Latent Embedding Techniques:} Latent embedding strategies have further bridged semantic representations and visual outcomes. CLIP-guided embeddings \cite{radford2021learningtransferablevisualmodels} map high-level descriptors (e.g., “impressionist sunset”) to diffusion processes, enabling seamless integration of textual and visual cues. Text-to-image diffusion models \cite{nichol2022glidephotorealisticimagegeneration} capitalize on these embeddings to generate visually coherent outputs from semantic prompts, enhancing creativity and personalization. Latent Alignment Models \cite{deng2018latentalignmentvariationalattention}  extend this concept by integrating alignment objectives to ensure style-specific embeddings maintain consistency with content structures. Normalizing Flow Diffusion (NFD) \cite{ho2020denoising} ensures superior alignment of latent distributions with target domains, reducing stylistic inconsistencies. Contrastive Latent Modeling \cite{mao2023contrastiveconditionallatentdiffusion} enhances diversity by introducing contrastive objectives, allowing diffusion models to generate richer stylistic variations without compromising content integrity.

\textbf{Cross-Domain Style Transfer Applications:} Diffusion models have showcased exceptional versatility in cross-domain style transfer, from sketch-to-painting transformations \cite{ramesh2022hierarchicaltextconditionalimagegeneration} to abstract-to-photorealistic synthesis. These methods utilize hierarchical representations and advanced embeddings to ensure robust style preservation across distinct artistic domains. Guided Multi-Modal Diffusion \cite{dhariwal2021diffusion} further extends their utility, enabling transformations between modalities like text, images, and videos.

\subsubsection{Optimized Training Processes for Stability and Scalability}
\label{sec:subsection_3.3.2}

Optimizing the training processes of diffusion models has been a critical focus in addressing stability, scalability, and computational efficiency. Numerous studies have introduced innovations in noise scheduling, reverse sampling, and hybrid frameworks to refine the training and application of diffusion models in style transfer tasks. These developments ensure that models generate high-quality outputs while efficiently handling complex artistic styles.

\textbf{Advanced Noise Schedules:} Noise scheduling is a core component of diffusion models, directly affecting convergence and output quality. Nichol and Dhariwal proposed cosine noise schedules \cite{nichol2021improveddenoisingdiffusionprobabilistic}, enhancing sampling efficiency and minimizing perceptual artifacts in style transfer outputs. \cite{anonymous2024improved} by increasing the sampling frequency of the signal-to-noise ratio logarithm, the model can focus on the key transition points between signal-dominant and noise-dominant, resulting in more robust and accurate predictions. Ho et al. \cite{ho2021cascadeddiffusionmodelshigh} developed simplified reverse diffusion pipelines, reducing the number of sampling steps while maintaining stylistic fidelity.

\textbf{Dynamic and Adaptive Sampling:}
Dynamic sampling techniques have emerged as a way to balance computational cost and output quality. AdaDiff \cite{tang2024adadiffacceleratingdiffusionmodels} dynamically adjusts computational resource allocation based on the importance of each step during the generation process. Step-Adaptive Training\cite{li2024stepsequalefficientgeneration} improves denoising capabilities in diffusion models by partitioning timesteps into distinct groups and fine-tuning the model for specialized denoising. AdaptiveDiffusion \cite{ye2024trainingfreeadaptivediffusionbounded} introduces a method to adaptively reduce the number of noise prediction steps in diffusion models, optimizing computational efficiency while maintaining output quality.

\begin{figure*}[h]    
    \centering
    \includegraphics[width=1\linewidth]{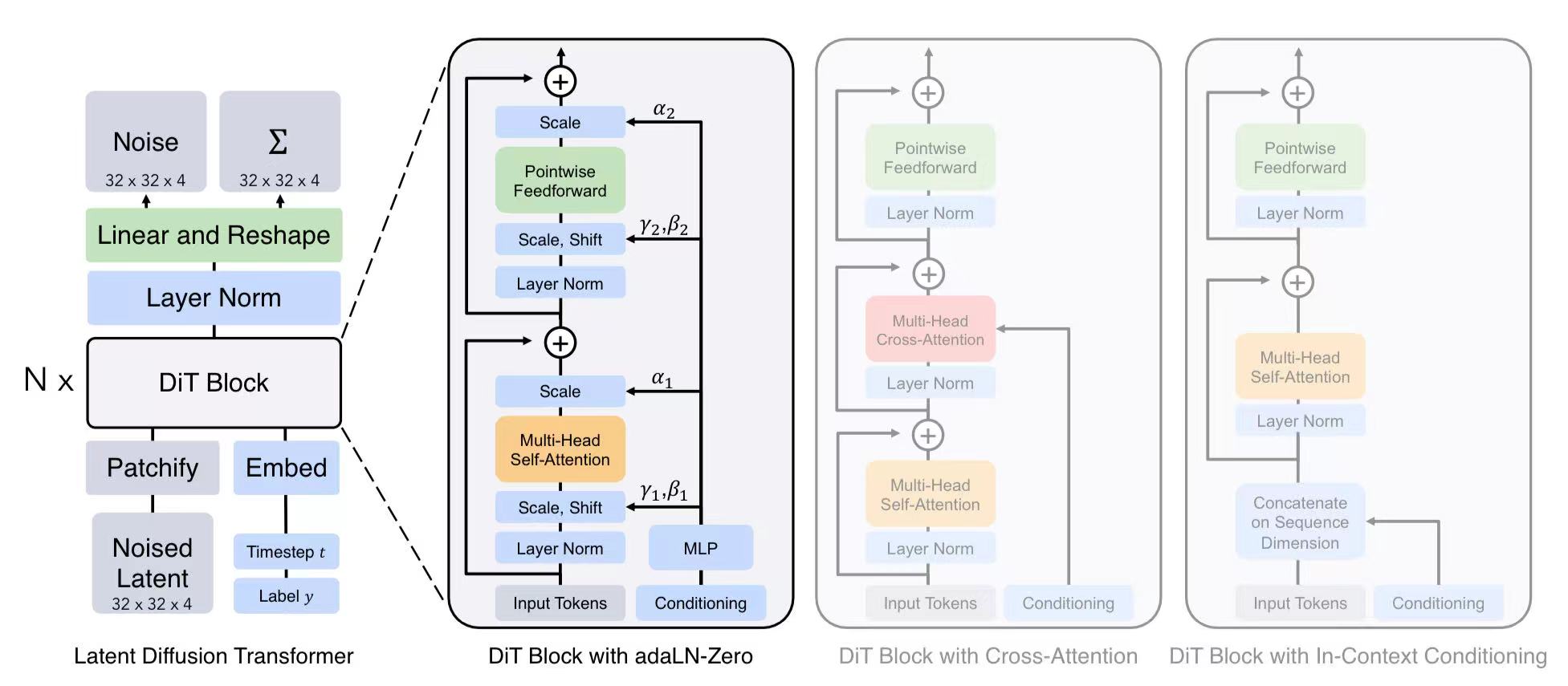} 
    \caption{The Diffusion Transformer (DiT) architecture.\cite{peebles2023scalablediffusionmodelstransformers}}
    \label{fig:DiT}
    \vspace{-0.4cm}
\end{figure*}

\textbf{Hybrid Training Frameworks:}
Integrating pre-trained models with diffusion architectures has proven effective in accelerating training. Masui et al. \cite{masui2024harnessinglatentdiffusionmodel} combined autoencoder-based latent representations with diffusion processes, creating a hybrid framework that enhances stability and detail preservation in style transfer. FastDiff 2 \cite{huang2023fastdiff} proposed initializing diffusion models with GAN pre-trained weights, significantly reducing training time and improving output quality. DIT \cite{peebles2023scalablediffusionmodelstransformers}, DiffiT \cite{hatamizadeh2024diffitdiffusionvisiontransformers}, U-ViT \cite{bao2023worthwordsvitbackbone}, and MDT \cite{gao2024mdtv2maskeddiffusiontransformer} extend these methods by using the transformer architecture as the feature encoder, as shown in the figure below \ref{fig:DiT}, to achieve a more scalable and expressive diffusion-based framework.

\textbf{Distributed Training and Scalability:}
Fokam et al. \cite{fokam2024asynchronousstochasticgradientdescent} proposed parallelizing the SGD updates in the layers of the model by updating them asynchronously from multiple threads. Sun et al. \cite{sun2024ecditscalingdiffusiontransformers} develop a new family of Mixture-of-Experts (MoE) models (EC-DIT) for diffusion transformers with expert-choice routing, optimized hardware utilization for complex art styles.

\textbf{Enhanced Training Objectives:} Choi et al. \cite{choi2022perceptionprioritizedtrainingdiffusion} achieve significant improvements in diffusion model performance by redesigning the weighting scheme of the objective function. Subsequently, Lin et al. \cite{lin2024diffusionmodelperceptualloss} use inductive biases to design better loss objectives and propose a novel self-perceptual loss that leverages the diffusion model itself as a perceptual loss.

\subsubsection{Advanced Conditioning Mechanisms for Multi-Modal and Cross-Domain Style Transfer}
\label{sec:subsection_3.3.3}

Diffusion models have emerged as powerful tools for multi-modal and cross-domain style transfer, offering unique capabilities for tasks that require transferring styles across diverse domains such as text-to-image, image-to-video, or even sketch-to-photograph. The innovations in conditioning mechanisms have made it possible to achieve more precise, flexible, and consistent transformations while maintaining high-quality outputs. 

\textbf{Attention-Driven Conditioning Mechanisms:}Attention mechanisms have significantly improved the ability of diffusion models to handle complex relationships between content and style. wu et al. \cite{wu2023harnessingspatialtemporalattentiondiffusion} propose a new text-to-image algorithm that adds explicit control over spatial-temporal cross-attention in diffusion models. Chen et al. \cite{chen2023trainingfreelayoutcontrolcrossattention} proposed another simple approach that achieves strong layout control without training or fine-tuning the image generator.

\textbf{Domain-Specific Conditioning for Cross-Domain Transfer:}Diffusion models have demonstrated exceptional capabilities in bridging domain gaps, with applications spanning style transfer, domain-specific transformations, and zero-shot domain adaptation. Their iterative refinement process, coupled with flexible latent space manipulations, enables high-fidelity image synthesis while maintaining stylistic and structural consistency. 
Cross-Domain Compositing with Pretrained Diffusion Models \cite{hachnochi2023crossdomaincompositingpretraineddiffusion} highlights the utility of pretrained backbones for compositional blending, allowing seamless integration of stylistic elements across domains without extensive retraining. By leveraging the pretrained capacity of diffusion models, this approach ensures fidelity in both style and structure, enabling versatile transformations such as embedding objects into new artistic contexts or adapting abstract textures to real-world imagery. The Domain-Guided Conditional Diffusion Model (DACDM) \cite{zhang2023domainguidedconditionaldiffusionmodel} introduces domain classifiers to guide diffusion sampling, enhancing unsupervised domain adaptation. This approach aligns latent representations with target domains, ensuring that transformations preserve domain-specific characteristics. Similarly, ZoDi \cite{azuma2024zodizeroshotdomainadaptation} addresses domain adaptation in scenarios lacking labeled data. By applying diffusion-based transformations, ZoDi achieves stylistic adaptation while maintaining structural coherence, facilitating tasks like sketch-to-photo or artistic-to-realistic rendering. Further extending the adaptability of diffusion models, Transfer Learning for Diffusion Models \cite{ouyang2024transferlearningdiffusionmodels} fine-tunes pretrained diffusion architectures for domain-specific applications, reducing data requirements and computational overhead.

\textbf{Cross-Modal Style Consistency:} Achieving robust cross-modal style consistency in text-guided image synthesis and editing involves ensuring that linguistic style cues align coherently with the corresponding visual attributes. This typically requires projecting both textual descriptions and reference imagery into a shared latent space that encodes semantic and stylistic information \cite{kwon2022clipstylerimagestyletransfer}\cite{rombach2022highresolutionimagesynthesislatent}. By employing refined cross-attention mechanisms and adaptive normalization layers during the generative diffusion process, models can smoothly inject style factors drawn from textual inputs,\cite{hertz2022prompttopromptimageeditingcross} thereby producing images whose textures, patterns, and artistic effects remain faithful to the original language-driven intent. Such frameworks often improve upon prior approaches by introducing more fine-grained control \cite{brooks2023instructpix2pixlearningfollowimage}: for instance, Prompt-to-Prompt editing methods allow the user to refine previously generated outputs with new textual instructions without sacrificing stylistic integrity. Moreover, additional conditioning modalities—such as edge maps, semantic layouts, or keypoints—can be integrated through architectures like ControlNet\cite{zhang2023addingconditionalcontroltexttoimage}, which helps maintain consistent style under complex, multimodal constraints. 
These architectures employ latent encoders or feature extraction modules to project heterogeneous modalities into a shared latent space\cite{choi2021ilvrconditioningmethoddenoising}, thereby enabling synergistic multi-path control. Hierarchical attention mechanisms and adaptive normalization techniques also dynamically inject diverse conditioning signals at various diffusion stages, ensuring that the generated images preserve underlying structural details while flexibly integrating a variety of stylistic attributes \cite{meng2022sdeditguidedimagesynthesis}. Empirical evidence demonstrates that such hybrid conditioning strategies notably surpass traditional single-condition paradigms in terms of stability, quality, and diversity, and offer enhanced versatility in text-guided style synthesis and user-customizable artistic style transfer scenarios \cite{saharia2022photorealistictexttoimagediffusionmodels}.

\textbf{Multi-Modal Embedding Strategies:} Multi-modal embedding strategies lie at the heart of current advancements in text-to-image synthesis, style transfer, and semantic alignment tasks. These strategies typically involve learning a joint representation space where textual and visual inputs can be mapped into semantically coherent embeddings, facilitating a direct correlation between linguistic descriptions and their corresponding visual features. By leveraging large-scale, diverse datasets and contrastive learning objectives, recent frameworks such as CLIP\cite{radford2021learningtransferablevisualmodels} and ALIGN\cite{jia2021scalingvisualvisionlanguagerepresentation} demonstrate a capacity to align heterogeneous modalities in a shared latent space, enabling zero-shot image recognition and flexible text-driven content generation \cite{ramesh2021zeroshottexttoimagegeneration}.

\subsubsection{Disentanglement and Content Preservation in Diffusion Models}
\label{sec:subsection_3.3.4}

\begin{figure}[h!]
    \centering
    \includegraphics[width=0.48\textwidth]{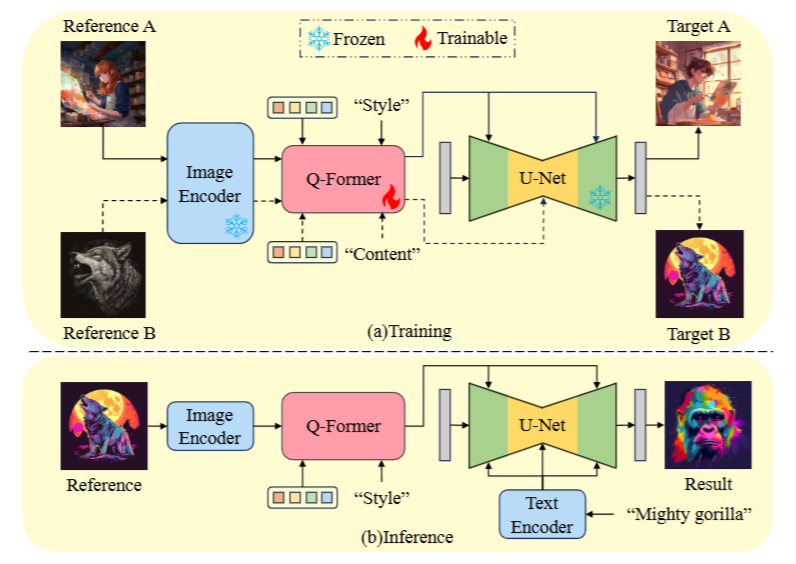}
    \caption{The training and inference paradigm of DEADiff\cite{qi2024deadiffefficientstylizationdiffusion}.}
    \label{fig:DEADiff}
\end{figure}

\textbf{Innovative Approaches to Disentanglement:}
Disentanglement in generative models aims to separate distinct factors of variation—such as style and content—to enable more controlled and interpretable image synthesis and manipulation. Recent advancements in diffusion-based models have introduced innovative techniques to achieve effective disentanglement, enhancing the flexibility and precision of style transfer applications. One prominent approach leverages conditional diffusion processes that explicitly factorize style and content by incorporating separate conditioning pathways for each attribute \cite{karras2021aliasfreegenerativeadversarialnetworks}. This allows the model to independently manipulate style features without altering the underlying content structure.  
Complementing these innovations, StyleDiffusion \cite{wang2023stylediffusioncontrollabledisentangledstyle} focuses on disentangled style representations. DRL\cite{jun2024disentanglingdisentangledrepresentationsimproved} designs dynamic Gaussian anchoring to enforce attribute separation of latent U-Net\cite{ronneberger2015unetconvolutionalnetworksbiomedical} and proposes Skip Dropout technology to modify u-net, promoting independence between latent units.
In addition, DEADiff\cite{qi2024deadiffefficientstylizationdiffusion} extracts disentangled feature representations from Q-Formers and injects them into mutually exclusive subsets of the cross-attention layer. The structure is shown in the figure below \ref{fig:DEADiff}, which achieves the best visual stylization effect and achieves the best balance between the text controllability inherent in the text-to-image model and the style similarity with the reference image.

\section{Evaluation Metrics and Key Innovations in Style Transfer}

\subsection{perceived distance}
\label{evaluation_add}

Researchers have long sought a "perceptual distance" metric that can measure the similarity of two images in a way that is highly consistent with human judgment. However, constructing such a perceptual metric is extremely challenging, mainly due to the following three fundamental problems:

\begin{enumerate}
    \item Reliance on high-order image structure
    \item Highly context-dependent
    \item Cannot strictly form a distance metric
\end{enumerate}

Take a simple example: when we compare a red circle with a red square and a blue circle, humans will have multiple similarity perceptions simultaneously. This context dependency makes it almost impossible to directly fit a function that conforms to human judgment.

\subsubsection{Perceptual Metrics}

Perceptual metrics are designed to simulate the human visual system's perception and judgment of image similarity. These metrics attempt to quantify the perceived distance between images and overcome the limitations of traditional pixel-level evaluation methods.

Peak Signal-to-Noise Ratio (PSNR)\cite{1284395}, as a traditional image quality evaluation metric, evaluates similarity by measuring the pixel-level error between the generated image and the original image. PSNR-HVS is an extension of PSNR that incorporates properties of the human visual system such as contrast perception. Although PSNR is simple to calculate and widely used, it does not fully reflect the human visual system's perception of image quality, so it often needs to be used in combination with other perceptual indicators in style transfer tasks. In contrast, Feature Similarity Index Measure (FSIM)\cite{5705575} can more accurately capture the structure and details in the image by utilizing the phase consistency and texture information of the image, thereby improving the ability to evaluate image similarity. 

The Structural Similarity Index (SSIM) \cite{wang2004image} is one of the earliest widely used indicators for image quality assessment. However, SSIM at a single scale may not capture the detailed information of the image at different scales. To solve this problem, MSSIM (Multi-Scale Structural Similarity Index) \cite{ABDELSALAMNASR2017399} was introduced, which can more comprehensively evaluate the structural similarity of images by calculating SSIM at multiple scales and performing weighted averaging. 

In addition, HDR-VDP (High Dynamic Range Visual Difference Prediction)\cite{mantiuk:2005:PredVisDiff} is specifically used to evaluate visual differences in high dynamic range images. HDR-VDP simulates the perception of image differences by the human visual system and can provide evaluation results that are more in line with human perception in high contrast and complex texture environments. Subsequently, HDR-VDP-2\cite{Mantiuk2011HDRVDP2AC}, HDR-VDP-3\cite{mantiuk2023hdrvdp3multimetricpredictingimage} was proposed, which provided better prediction results for visibility than the original HDR-VDP and VDP indicators.

\subsubsection{Content retention assessment}

In the multi-dimensional evaluation of style transfer, the evaluation of content preservation is particularly critical, focusing on the similarity between the generated image and the original content image.

CLIP Score\cite{hessel2022clipscorereferencefreeevaluationmetric} uses the advantages of the CLIP\cite{radford2021learningtransferablevisualmodels} model in visual and language understanding to calculate the similarity between the generated image and the original content image in the semantic space to evaluate the content retention. In contrast, Content Preservation Loss (CPL)\cite{kang2023contentpreservingimagetranslation} uses L2 loss to calculate pixel-level differences, or extracts features through convolutional neural networks and compares the feature differences between the two images to measure the degree of content information retention. Although CPL usually focuses on more fine-grained content preservation evaluation, its limitation is that over-emphasizing the precise matching of pixels or features may limit the style diversity of images.

\subsubsection{Diversity Metrics}

These metrics focus on the diversity, breadth of distribution, and degree of information variation of generated images. We collectively refer to them as "Diversity Metrics."

Diversity Metrics include Entropy (EN), Correlation Coefficient (CC), Standard Deviation (SD), Spectral Convergence Divergence (SCD), Mutual Information (MI), and Fuzzy Mutual Information (FMI). Among them, EN measures the information entropy of the image. A higher entropy value indicates a greater complexity of the image content, which is usually associated with higher diversity; CC is used to evaluate the correlation between image pixels. A lower correlation may mean greater diversity of generated images; SD reflects the diversity of images by quantifying the changes in image pixel values. A larger standard deviation indicates a greater change in image content; SCD evaluates the diversity of generated images through spectral differences, which can reveal whether the model has mode collapse; MI and FMI judge the richness of the generated image content by measuring the degree of information sharing in the image. Higher mutual information means higher information retention and diversity.

\subsubsection{Deep learning related indicators}

In the evaluation of generative models, the Fréchet Inception Distance (FID)\cite{heusel2018ganstrainedtimescaleupdate}  has emerged as a pivotal metric for assessing both the fidelity and diversity of generated images. Initially introduced by Heusel et al., FID \cite{heusel2018ganstrainedtimescaleupdate} measures the distance between the feature distributions of real and generated images, leveraging the Inception network's representations to provide a quantitative assessment of image quality and variability. However, as the field has advanced, researchers identified limitations in the traditional FID approach, particularly concerning its sensitivity to the underlying data distribution and its ability to capture nuanced image details.

To address these shortcomings, subsequent studies have proposed enhancements and alternative metrics aimed at refining the evaluation process. For instance, Reliable Fidelity and Diversity Metrics for Generative Models\cite{naeem2020reliablefidelitydiversitymetrics} expanded the evaluation framework by integrating multiple metrics alongside FID to provide a more comprehensive assessment of generative models' performance. Further advancements include the introduction of Compound Fréchet Inception Distance for Quality Assessment of GAN Created Images\cite{nunn2021compoundfrechetinceptiondistance}, which combines FID with additional feature-based metrics to enhance sensitivity to various aspects of image quality, thereby offering a more robust evaluation tool. Additionally, the clean-FID\cite{parmar2022aliasedresizingsurprisingsubtleties} metric was developed to mitigate noise and irrelevant information in the feature representations, thereby increasing the robustness and accuracy of FID calculations. In Rethinking FID\cite{jayasumana2024rethinkingfidbetterevaluation}, researchers critically examined the theoretical underpinnings and practical applications of FID, proposing modifications to address identified biases and improve its applicability across diverse generative tasks. Furthermore, Learning Perceptual Image Patch Similarity, as a perceptual similarity metric based on deep features, can better reflect the similarity of human perception of images by comparing the features extracted from two images at different levels of deep neural networks. Building on this foundation, the Robust Learning Perceptual Image Patch Similarity (R-LPIPS)\cite{ghazanfari2023rlpipsadversariallyrobustperceptual} indicator is proposed, which is a new indicator that uses adversarial training deep features.

\subsection{Key Techniques and Innovations}
\label{app:keytech}

\subsubsection{Artistic Merit}

Artistic merit stands as a cornerstone in style transfer research, as it directly determines the aesthetic quality and visual appeal of the generated images. The significance of artistic merit in style transfer is twofold: first, it ensures the preservation of the original artistic elements that make the style unique and recognizable; second, it guarantees the naturalness and visual coherence of the stylized results. This becomes particularly crucial as the field moves towards more sophisticated applications in digital art creation, content generation, and creative industries.

The ability to capture and transfer artistic style in a meaningful way remains a fundamental challenge in style transfer. Early methods like Neural Style Transfer (NST) \cite{gatys2015neural} established the foundation by separating content and style representations through CNN features. Subsequent innovations have focused on preserving artistic integrity while transferring style. AdaIN \cite{huang2017arbitrary} introduced adaptive instance normalization to better align content and style features, while Avatar-Net \cite{sheng2018avatarnetmultiscalezeroshotstyle} developed a multi-scale approach to capture artistic elements at various levels of abstraction.

Recent advancements have introduced more sophisticated approaches to enhance artistic quality. DiffStyler \cite{li2024diffstyler} proposes a novel framework that leverages diffusion models to achieve more precise style control while maintaining artistic coherence. The method introduces a style-aware guidance mechanism that helps preserve the intricate details and artistic nuances of the style reference. CreativeSynth \cite{huang2024creativesynthcreativeblendingsynthesis} advances this further by introducing a synthesis mechanism that can decompose and recombine artistic elements in novel ways, allowing for more creative and diverse stylization results while maintaining artistic fidelity.

More recent approaches like CLIPstyler \cite{kwon2022clipstylerimagestyletransfer} have leveraged text-guided mechanisms to provide more precise control over artistic elements, enabling users to specify desired artistic qualities through natural language descriptions. This text-driven approach has opened new possibilities for fine-grained artistic control and customization in style transfer applications.

\begin{figure}[h]
    \centering
    \includegraphics[width=0.48\textwidth]{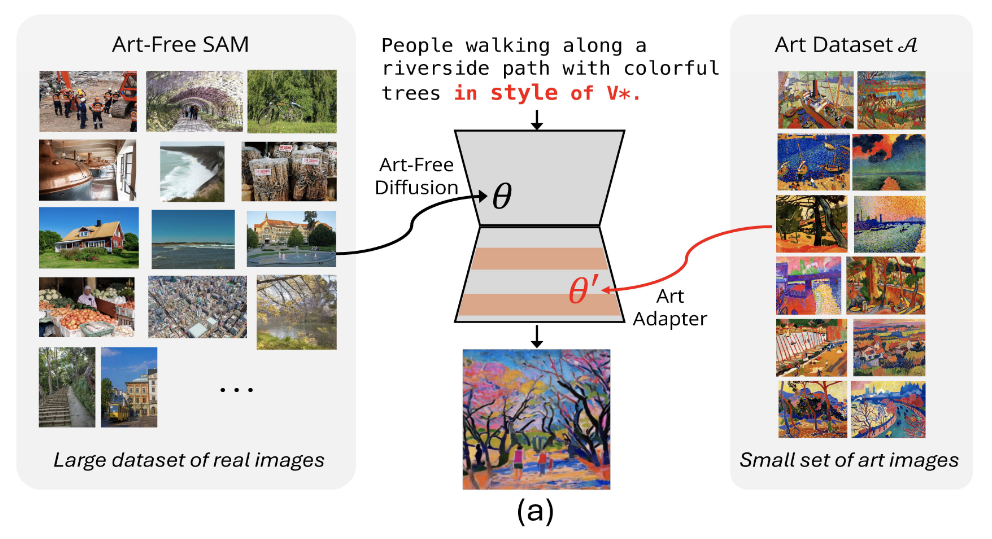} 
    \caption{Overview of the Art-Free framework\cite{ren2024art-free}. The model consists of three main components: (left) Art-Free SAM trained on a large dataset of real-world images without artistic content, (middle) an Art-Free Diffusion model that processes text prompts and generates images, and (right) a specialized Art Adapter trained on a small set of artistic images. The framework enables artistic style transfer by combining general image understanding from non-artistic data with targeted style adaptation, demonstrating that high-quality artistic generation can be achieved without extensive art-specific training data.}
    \label{fig:artfree_framework}
\end{figure}

A noteworthy recent development in this field is the Art-Free Generative Models approach \cite{ren2024art-free}, which challenges the conventional wisdom about the necessity of extensive artistic training data. This innovative method demonstrates that high-quality artistic style transfer can be achieved without training on art-specific datasets, instead utilizing an "art adapter" that learns artistic styles from minimal examples. The structure is shown in the figure below \ref{fig:artfree_framework}. This approach not only advances the technical capabilities of style transfer but also addresses ethical concerns regarding artistic copyright and data usage. Furthermore, it points to a promising direction where style transfer systems can become more efficient and accessible while maintaining high artistic merit.

\subsubsection{Visual Quality}

Visual quality serves as a critical metric in style transfer evaluation, as it directly impacts the practical applicability and user acceptance of stylized results. High visual quality ensures that the stylized images not only capture the desired artistic effects but also maintain the essential characteristics of the content image, such as structural integrity, texture consistency, and detail preservation. This becomes particularly crucial in professional applications like digital content creation, where the output quality directly affects user engagement and commercial value.

Visual quality has been significantly enhanced through several key innovations in the field. The introduction of perceptual loss by Johnson et al. \cite{johnson2016perceptual} marked a crucial advancement in maintaining high-quality results while enabling real-time processing. Histogram matching techniques \cite{risser2017histogram} have improved color consistency and overall visual coherence. SEAN \cite{zhu2020sean} further advanced visual quality by enabling precise control over different image regions, allowing for more refined and visually appealing results.

Portrait Diffusion \cite{liu2023portrait} introduces a specialized framework for portrait style transfer, addressing the unique challenges in maintaining facial features and expressions while applying artistic styles. The method employs a novel quality-aware attention mechanism that helps preserve crucial facial details and ensures natural-looking results. FreeStyle \cite{he2024freestyle} advances the field by introducing a flexible architecture that can maintain high visual fidelity across diverse style types, particularly excelling in preserving fine details and reducing artifacts in complex scenes.

On the other hand, researchers have placed significant emphasis on preserving structural integrity when applying artistic styles. For instance, ArtFlow \cite{an2021artflowunbiasedimagestyle} has demonstrated remarkable success in maintaining the content structure during style transfer, ensuring that the essential geometric and spatial relationships of the original image are retained. This focus on structural preservation is crucial for applications where the accuracy of the content's layout and form is paramount, such as in architectural visualization and technical illustrations. Additionally, methods like Deep Image Prior \cite{Ulyanov_2020} and Neural Style Transfer with Structure Preservation \cite{TIP20_SP_NPR} have introduced innovative techniques to balance artistic transformation with the fidelity of structural elements. These advancements highlight the field's dedication to achieving a harmonious blend of artistic creativity and structural accuracy, enabling more reliable and versatile applications in various professional domains.

\subsubsection{Computational Efficiency}

Computational efficiency represents a critical aspect of style transfer research, as it directly impacts the practical applicability of these methods in real-world scenarios. The balance between processing speed and resource consumption becomes particularly crucial when deploying style transfer systems in resource-constrained environments such as mobile devices or web applications.

The evolution of style transfer methods has been marked by significant improvements in computational efficiency. The initial optimization-based approach by Gatys et al. \cite{gatys2015neural} required several minutes to process a single image. The transition from optimization-based approaches to feed-forward networks, pioneered by Johnson et al. \cite{johnson2016perceptual} and Ulyanov et al. \cite{ulyanov2016texture}, dramatically reduced processing time from minutes to milliseconds, marking a significant milestone in efficiency improvement.

Subsequent research focused on architectural optimization. Light Transfer \cite{ulyanov2017improved} specifically addressed mobile deployment challenges by developing lightweight architectures, achieving significant reduction in model size while maintaining reasonable quality. FastPhotoStyle \cite{li2018closedformsolutionphotorealisticimage} introduced a two-stage pipeline that separated style transfer into content preservation and style enhancement stages, enabling faster processing while maintaining high-quality results.

In addition to these innovations, recent methods have specifically targeted computational efficiency. For example, Fast Neural Style Transfer via Random Fourier Features \cite{tancik2020fourierfeaturesletnetworks} proposed using randomized Fourier features to accelerate the computation of style transfer without sacrificing visual quality. This technique effectively reduces the complexity of matrix operations, leading to faster execution times. Furthermore, the work of Zheng et al. \cite{Zheng_2024_CVPR} introduced Puff-Net, an efficient style transfer method that utilizes a pure content and style feature fusion network. This approach enhances computational efficiency by effectively separating content and style information, allowing for faster processing of high-resolution images while preserving high-quality outputs. Puff-Net significantly reduces computational overhead while maintaining impressive visual fidelity, making it suitable for real-time applications and mobile devices.

Modern approaches have continued this trend towards efficiency, with methods like AdaAttN \cite{liu2021adaattnrevisitattentionmechanism} achieving efficient attention-based transfer while maintaining quality through innovative attention mechanisms that reduce computational complexity. These improvements have made style transfer increasingly practical for real-world applications and mobile devices, enabling applications ranging from real-time video processing to mobile style transfer apps.

\subsubsection{Robustness and Generalization}

Robustness and generalization capabilities have emerged as crucial factors in style transfer research, as they determine the reliability and versatility of these systems across diverse real-world scenarios. These aspects become particularly important when style transfer methods need to handle varying image conditions, different artistic styles, and complex content structures while maintaining consistent quality.

Robustness and generalization capabilities have seen significant evolution in recent years. Universal Style Transfer methods \cite{li2017universal} marked a significant advancement by enabling transfer of arbitrary styles without retraining, establishing a foundation for more flexible applications. This was achieved through adaptive instance normalization techniques that could better handle style variations. Following this, StyleBank \cite{chen2017stylebankexplicitrepresentationneural} introduced a modular approach that improved robustness by learning separate style-specific filters, allowing for better adaptation to different style characteristics.

Recent approaches have focused on improving stability across different content-style pairs and handling challenging cases. Methods like ArtFlow \cite{an2021artflowunbiasedimagestyle} have addressed this through invertible neural networks, which provide more stable and reversible style transformations. SANet \cite{fan2017sanetstructureawarenetworkvisual} introduced self-attention mechanisms to better handle spatial dependencies, significantly improving the handling of complex structural patterns and extreme style variations.

The development of more sophisticated normalization techniques and the integration of multi-scale processing have contributed to more robust results across various scenarios. AdaIN \cite{huang2017arbitrary} advanced this further by introducing adaptive instance normalization that could better handle style variations while maintaining content integrity. These improvements have made style transfer systems more reliable and adaptable to different use cases, from artistic creation to commercial applications.

\subsubsection{Multimodal Capability}

The latest developments in style transfer have expanded beyond purely visual domains to incorporate multiple modalities. Modern systems can process and transfer styles across different types of data, including text, audio, and video. AdaIN's extension to multimodal transfer \cite{huang2018multimodal} demonstrated the possibility of diverse style outputs from a single content input. Recent text-guided approaches like CLIPstyler have further expanded these capabilities by enabling natural language control over style transfer. These multimodal capabilities have opened new possibilities for creative applications and cross-domain style transfer, though challenges remain in achieving consistent quality across different modalities.

\subsubsection{Ethical Considerations and Safety}

The widespread adoption of AI-powered style transfer has raised critical ethical concerns, spanning intellectual property rights, potential misuse, privacy, and cultural sensitivity. Addressing these issues requires a balanced approach that fosters innovation while ensuring responsible deployment.

A primary ethical challenge is copyright and intellectual property rights. Style transfer systems can replicate distinctive artistic styles, raising concerns about attribution and fair compensation, particularly when commercial applications use protected artworks without authorization. This issue has sparked discussions on frameworks that respect artists’ rights while encouraging computational creativity.

Another significant concern is the potential misuse of style transfer to create deceptive or misleading content. These techniques can generate fake images or videos that mimic authentic works, intersecting with broader issues such as deepfakes and digital manipulation. To counteract this, researchers have proposed watermarking techniques and style transfer signatures to improve transparency and traceability.

Privacy risks also arise when style transfer is applied to personal images or sensitive content, particularly in social media and mobile applications. The ability to modify identifying features in images raises concerns about data protection and user consent, necessitating stricter guidelines on how these technologies handle personal data.

Cultural sensitivity is another key issue. Style transfer models that apply culturally significant artistic styles must consider historical and contextual significance to avoid cultural appropriation. Ensuring the respectful use of traditional artistic elements is essential to maintaining cultural integrity and representation.

From a safety perspective, addressing technical vulnerabilities is crucial. This includes improving system robustness against adversarial attacks, preventing the generation of harmful content, and implementing content filtering mechanisms. Some research efforts focus on integrating built-in safety checks and ethical constraints into style transfer frameworks.

To mitigate these challenges, the research community is developing ethical frameworks and best practices. These include guidelines for proper attribution, transparent disclosure of AI-generated content, and mechanisms allowing artists to opt out of dataset inclusion. Additionally, the emphasis on interpretable and controllable style transfer systems aims to enhance oversight and accountability in real-world applications.

\section{Applications and Practices of Style Transfer}
\label{app:domains}
\subsection{Portraits-Style-Transfer}
\label{sec:portraits}
As an important application field of image style transfer, portrait style transfer aims to apply a specific artistic style to portrait images with high quality, and realize the transformation from real photos to works of art. In recent years, with the rapid development of generative adversarial networks, diffusion models and other deep learning technologies, portrait style transfer technology has made significant progress in generation quality, style diversity and application breadth. In particular, in the generation of two-dimensional (2D) and cartoon-style portraits, researchers have proposed many innovative methods further to enrich the expressiveness and practicality of portrait style transfer.

Early research in portrait style transfer primarily leveraged Generative Adversarial Networks (GANs) to achieve style transformation of portrait images through adversarial training. Yi et al. proposed APDrawingGAN \cite{YiLLR19}, which utilizes GANs to convert portrait photographs into pencil sketch styles, demonstrating high-quality line details and realism. This approach laid the foundation for subsequent cartoonized portrait generation. Following this, Shi et al. introduced WarpGAN \cite{shi2019warpganautomaticcaricaturegeneration}, which incorporates image geometric warping techniques to automatically generate cartoonized portraits with exaggerated features, further enriching the expressive capabilities of style transfer. Building on this, Back et al. fine-tuned StyleGAN2 to develop Cartoon-StyleGAN \cite{back2021finetuningstylegan2cartoonface}, focusing on the generation of cartoon-style facial images and significantly enhancing the detail and realism of the generated images. Additionally, StyleCariGAN \cite{Jang2021StyleCari} and CariMe \cite{gu2021carime} achieved multi-style blended portrait generation by integrating multiple style generation modules, offering users a diverse range of artistic expressions.

As technology has continued to evolve, researchers have begun exploring more dynamic and diverse generation methods. Kim et al. introduced DynaGAN \cite{Kim2022DynaGAN}, which dynamically adjusts the structure of the generative network to adapt to different portrait style requirements, thereby enhancing the model's flexibility and adaptability. BlendGAN \cite{liu2021blendgan}, on the other hand, merges multiple style generation modules to achieve multi-style blended portrait generation, further increasing the diversity and artistic quality of the generated images.

In addition to GANs, diffusion models have emerged as a significant method for portrait style transfer in recent years. Liu et al. developed PortraitDiffusion \cite{liu2023portrait}, which leverages the powerful generative capabilities of diffusion models to achieve high-fidelity portrait style transfer while preserving the details and structure of the original images. These methods typically offer better stability and higher generation quality, making them suitable for high-quality artistic portrait transformations. Furthermore, Chefer et al. introduced TargetCLIP \cite{chefer2021targetclip}, which combines the CLIP model to enable text-guided style transfer, allowing for more flexible and controllable image generation and expanding the application boundaries of style transfer.

To enhance the detail representation and adaptability of style transfer models, researchers have proposed various innovative methods. For instance, Men et al. presented DCT-Net \cite{men2022dct}, which utilizes the Discrete Cosine Transform (DCT) to capture the frequency domain features of images, effectively enhancing the detail representation in style transfer. IP-Adapter \cite{ye2023ip-adapter} and SSR-Encoder \cite{zhang2024ssr} improve the encoder structure and adapter modules, respectively, to enhance the model's adaptability and generalization across different styles, further increasing the practicality and flexibility of style transfer models.

Real-time style transfer is another active area of research. Wang et al. developed InstantStyle \cite{wang2024instantstyle}, which focuses on optimizing the computational efficiency of the model to achieve real-time portrait style conversion, particularly suitable for resource-constrained environments such as mobile devices. Concurrently, Liu et al. introduced ZePo \cite{liu2024zepo}, which employs zero-shot pose optimization to enable portrait style transfer under varying poses, thereby enhancing the model's practicality and diversity.

In practical applications of portrait style transfer, these technologies are widely used in the creation of personalized artistic works and automated drawing. For example, Gao et al.'s AiSketcher \cite{gao2020makingrobotsdrawvivid} demonstrates the capability of robots to autonomously draw vivid portraits, showcasing the potential of portrait style transfer in automated artistic creation. CreativeSynth \cite{huang2024creativesynth} combines multiple generative techniques to provide a creative synthesis platform that supports users in generating diverse artistic portraits. Additionally, Jones et al.'s Pair Customization \cite{jones2024customizingtexttoimagemodelssingle} allows users to customize style transfer models according to personal preferences, further expanding the application boundaries of portrait style transfer.

\subsection{Video Style Tranfer}
\label{sec:video}

Video style transfer, as a significant extension of image style transfer, aims to apply specific artistic styles to video sequences with high quality, thereby transforming real-world videos into artistic renditions. Early studies primarily relied on Generative Adversarial Networks and linear transformation methods, performing style transfer on individual frames followed by post-processing techniques to ensure temporal consistency. Gao et al. proposed ReCoNet \cite{gao2018reconetrealtimecoherentvideo}, which utilizes a real-time coherence network to perform style transfer on each frame of a video sequence while maintaining temporal coherence. Li et al. introduced Learning Linear Transformations \cite{li2018learning}, which simplifies the computational complexity of video style transfer by learning linear transformation matrices, thereby improving processing efficiency.

As research has advanced, more complex and efficient methods have been proposed to address the challenges of temporal consistency and style diversity in video style transfer. Yang et al. developed VToonify \cite{yang2022Vtoonify}, which combines multi-scale feature extraction and style fusion mechanisms to achieve high-quality cartoonized video generation. Liu et al. introduced AdaAttN \cite{liu2021adaattnrevisitattentionmechanism}, which revisits the attention mechanism by enhancing detail representation and overall consistency in style transfer through adaptive attention modules. Kasten et al. proposed Layered Neural Atlases \cite{kasten2021layeredneuralatlasesconsistent}, adopting a layered neural atlas approach to effectively separate and reconstruct different hierarchical structures within videos, thereby enhancing the precision and stability of style transfer. Wu et al. introduced CCPL \cite{wu2022ccpl}, which incorporates a content similarity preservation loss to further enhance the fidelity of content during the video style transfer process.

In recent years, the field of video style transfer has seen a surge of innovative methods, further advancing the technology. Qi et al. proposed FateZero \cite{qi2023fatezerofusingattentionszeroshot}, which achieves high-quality video style transfer in scenarios with limited large-scale training data by integrating attention mechanisms and zero-shot learning. Wen et al. developed CAP-VSTNet \cite{wen2023capvstnetcontentaffinitypreserved}, which optimizes content preservation during the style transfer process through a Content Affinity Preserved network. Yang et al. introduced Rerender A Video \cite{yang2023rerender} and Huang et al. presented Style A Video \cite{huang2023style}, which employ re-rendering techniques and style adaptation mechanisms, respectively, to enable more flexible and diverse video style transformations. Additionally, Chen et al. proposed Control A Video \cite{chen2023controlavideo}, which incorporates control mechanisms allowing users to perform more precise control and adjustments during the video style transfer process.

Moreover, Sanakoyeu et al. proposed Adaptive Style Transfer \cite{sanakoyeu2018styleawarecontentlossrealtime}, which enhances the effectiveness and efficiency of real-time video style transfer by incorporating an adaptive, style-aware content loss. Recently, Xu et al. and Cui et al. introduced the Hallo \cite{xu2024hallo}, Hallo2 \cite{cui2024hallo2}, and Hallo3 \cite{cui2024hallo3} series of methods, which utilize hierarchical style transfer strategies and optimization algorithms to further improve the quality and application scope of video style transfer.

In practical applications, video style transfer technologies are widely utilized in fields such as film production, virtual reality, game development, and social media content creation. For instance, VToonify \cite{yang2022Vtoonify} has been extensively applied in character design for animation and games, enhancing the efficiency and diversity of artistic creation. CreativeSynth \cite{huang2024creativesynth} combines multiple generative techniques to provide a creative synthesis platform that supports users in generating diverse video artistic works. Additionally, the application of the Hallo series methods \cite{xu2024hallo,cui2024hallo2,cui2024hallo3} further expands the boundaries of video style transfer, enabling users to apply style transfer technologies across various scenarios and requirements flexibly.

\subsection{3D Style Transfer}
\label{sec:3d}
3D style transfer, as a significant extension of image style transfer, aims to apply specific artistic styles to 3D models and animations with high quality, transforming real or standard 3D models into artistic 3D renditions. In recent years, advancements in deep learning and computer graphics technologies have led to significant progress in 3D style transfer, particularly in terms of generation quality, style diversity, and application breadth. This progress is especially evident in fields such as 3D animation production, virtual reality, and game development, showcasing a broad range of potential applications.

Early research in 3D style transfer primarily relied on Generative Adversarial Networks (GANs) and linear transformation methods. These approaches involved performing style transfer on individual frames of 3D models or animations, followed by post-processing techniques to ensure spatial and temporal consistency. For instance, RSMT\cite{10.1145/3588432.3591514} utilizes a real-time coherence network to perform style transfer on each frame of a 3D model while maintaining spatial coherence. Similarly, Li et al. introduced Learning Linear Transformations\cite{li2018learning}, which simplifies the computational complexity of 3D style transfer by learning linear transformation matrices, thereby improving processing efficiency and providing a foundation for subsequent research.

\begin{figure*}[h]    
    \centering
    \includegraphics[width=1\linewidth]{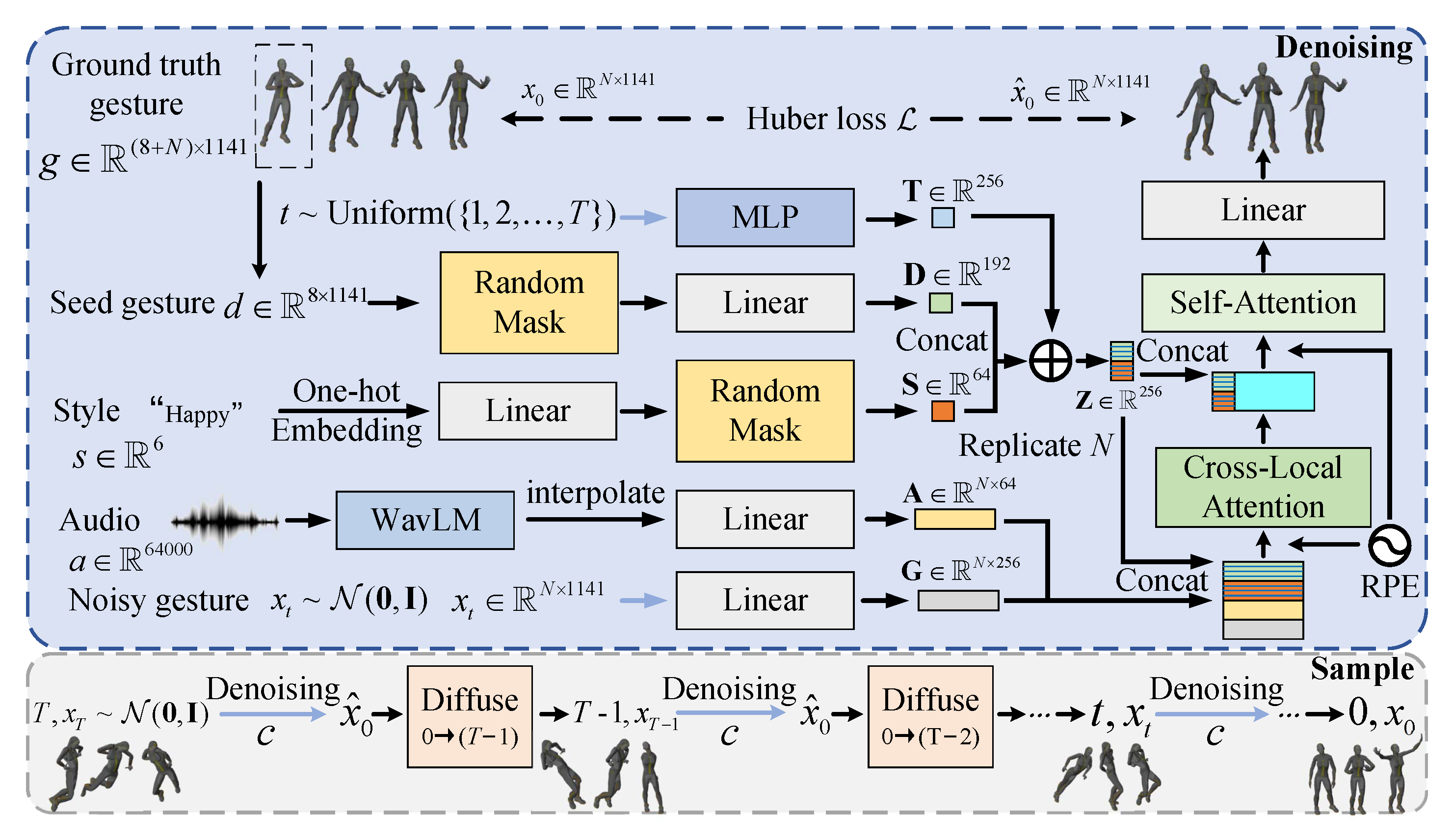} 
    \caption{The figure illustrates a diffusion-based framework for gesture generation, as proposed in DiffuseStyleGesture \cite{ijcai2023p650}. Inputs, including ground truth gestures, seed gestures, style labels, and audio features, are processed through random masking, feature extraction, and cross-local attention modules to generate gesture representations. The model iteratively denoises and refines gestures through the diffusion process, optimizing predictions with the Huber loss to produce realistic gesture sequences.}
    \label{fig:DiffuseStyleGesture.png}
    \vspace{-0.4cm}
\end{figure*}

As research has progressed, more complex and efficient methods have been proposed to address challenges related to spatial consistency and style diversity in 3D style transfer. DiffuseStyleGesture\cite{ijcai2023p650} combines multi-scale feature extraction and style fusion mechanisms to achieve high-quality gesture animation style transfer, as illustrated in Figure \ref{fig:DiffuseStyleGesture.png}. CAMDM\cite{camdm} enhances the detail representation and overall consistency of 3D style transfer through an improved attention mechanism. Additionally, Mason et al. introduced Local Motion Phases\cite{mason2022local}, which leverages local motion phase information to effectively maintain the fluidity and naturalness of animated motions.

To evaluate and advance 3D style transfer technologies, researchers have developed various datasets. For example, the 100STYLE\cite{mason2022local} dataset comprises four million frames of stylized motion capture data, specifically designed for 3D style transfer research. The Bandai-Namco-Research\cite{kobayashi2023motion} dataset includes a total of 36,673 frames, covering a wide range of styles in 3D models and animations, thereby further supporting the development and evaluation of 3D style transfer methods.

In practical applications, 3D style transfer technologies are widely utilized in 3D animation production, virtual reality, game development, and social media content creation. For instance, RSMT\cite{10.1145/3588432.3591514} has been applied in 3D animation and virtual character design, enhancing the efficiency and diversity of artistic creation. DiffuseStyleGesture\cite{ijcai2023p650} offers a creative synthesis platform by integrating multiple generative techniques, enabling users to generate diverse 3D animation artworks. Furthermore, the applications of CAMDM\cite{camdm} and Local Motion Phases\cite{mason2022local} have expanded the boundaries of 3D style transfer, allowing users to apply style transfer technologies across various scenarios and requirements flexibly.

\subsection{Text Style Transfer}
\label{sec:text}
Text style transfer aims to transform text from source style to target style while preserving its semantic content. Existing approaches can be broadly categorized into supervised methods based on parallel corpora and unsupervised methods using non-parallel data. 

Supervised approaches employ sequence-to-sequence models to directly learn style transformation mappings, though they are limited by the scarcity of parallel data. Unsupervised methods, which are more prevalent, achieve transfer primarily through content and style disentanglement: disentangled representation methods encode text into separate content and style latent representations before recombining them with target style features; adversarial methods incorporate discriminators to ensure style accuracy; and attribute control methods enable controllable transfer by explicitly modeling style attributes in the generation model.

Early research focused on supervised learning with parallel corpora, exemplified by the Sequence to Better Sequence model \cite{pmlr-v70-mueller17a}. However, due to parallel data scarcity, the field shifted towards unsupervised approaches. Shen et al. pioneered non-parallel style transfer using adversarial training \cite{shen2017styletransfernonparalleltext}. This was followed by various disentanglement-based approaches: ARAE employed adversarially regularized autoencoders \cite{2017arXiv170604223J}; Fu et al. explored multiple disentanglement strategies \cite{fu2018style}; and the Delete, Retrieve, Generate framework introduced explicit content preservation \cite{li2018deleteretrievegeneratesimple}.
Content preservation has seen significant advances: Tian et al. introduced dedicated content constraints \cite{tian2018structured}; John et al. achieved precise feature disentanglement \cite{john2018disentangled}; and SHAPED enhanced semantic consistency through hierarchical alignment \cite{zhang2018shaped}. Generation quality improvements came through language model discriminators \cite{yang2018unsupervised} and the Style Transformer architecture \cite{dai2019style}. Specialized applications emerged, including politeness transfer \cite{madaan2020politenesstransfertaggenerate} and profanity redaction \cite{tran2020friendlyonlinecommunityunsupervised}.
Recent trends show development in multiple directions: multi-attribute controllable approaches, exemplified by Lample et al.'s work \cite{lample2019multiple} and SETTP \cite{jin2024settpstyleextractiontunable}; integration of pre-trained language models, as in PowerTransformer \cite{ma2020powertransformerunsupervisedcontrollablerevision} and BERT-based methods \cite{hayati2021doesbertlearnhumans}; and improved evaluation methodologies \cite{pang1810learning}. Despite progress, challenges remain in content preservation, style accuracy, generation quality, and generalization capability.
The field of text style transfer continues to evolve, with future work likely focusing on addressing these challenges while expanding applicability across domains and languages. The integration of advanced pre-trained models and robust evaluation metrics will be crucial for future developments in this dynamic research area.

\subsection{Domain Adaptation}
\label{sec:domain}
Domain adaptation addresses the challenge of transferring knowledge learned from a source domain to a target domain where the data distributions differ. In the context of style transfer applications, domain adaptation techniques are particularly relevant for bridging the gap between synthetic and real domains \cite{liu2020stereoganbridgingsynthetictorealdomain}. Existing approaches can be categorized into supervised methods requiring paired domain data and unsupervised methods utilizing unpaired data. While supervised approaches directly learn domain mappings through paired training, they are limited by the availability of cross-domain paired data. Unsupervised methods, which are more common, achieve adaptation through various techniques including: feature alignment methods that align domain distributions in a shared feature space; adversarial learning approaches that employ discriminators to reduce domain discrepancy; and conditional generation methods that explicitly model domain-specific attributes \cite{murez2017imageimagetranslationdomain, 8578243}.

Early work in this field focused on supervised learning approaches with paired domain data. However, the difficulty in obtaining paired cross-domain data led to increased interest in unsupervised methods. Li et al. pioneered cycle-consistent adversarial transfer \cite{li2019cycleconsistentconditionaladversarialtransfer}, while others explored different adaptation strategies: GA-DAN proposed geometry-aware domain adaptation \cite{zhan2019gadangeometryawaredomainadaptation}; PointDAN introduced multi-scale adaptation for 3D point clouds \cite{qin2019pointdanmultiscale3ddomain}; and GCAN developed graph-based conditional adaptation networks \cite{8953825}.

Recent advances have focused on improving adaptation quality through various means: Yang et al. developed one-shot adaptation for face generation \cite{yang2020oneshotdomainadaptationface}; Long et al. proposed joint adaptation networks \cite{long2017deeptransferlearningjoint}; and Li et al. introduced domain-conditioned adaptation networks \cite{li2020domainconditionedadaptationnetwork}. Applications have expanded to include specialized tasks such as image dehazing \cite{9156566}, cross-domain document detection \cite{li2020crossdomaindocumentobjectdetection}, and animal pose estimation \cite{cao2019crossdomainadaptationanimalpose}.

Current research trends emphasize improving adaptation efficiency \cite{yu2019acceleratingdeepunsuperviseddomain}, handling domain-specific challenges \cite{9156713}, and developing more robust feature representations \cite{huang2020probabilityweightedcompactfeature}. Despite progress, challenges remain in achieving stable adaptation, maintaining semantic consistency, and generalizing across diverse domains.

\section{Datasets and Evaluation Criteria}

\subsection{Construction and Analysis of Datasets}
\label{Datasets}

In the domain of anime, the Danbooru series datasets (including Danbooru2017\cite{danbooru2018}, Danbooru2018\cite{danbooru2018},  Danbooru2020\cite{danbooru2020}, Danbooru2021\cite{danbooru2021}) have emerged as pivotal resources for anime style transfer research due to their extensive image collections and rich tag annotations. These datasets comprise millions of high-quality anime images annotated with hundreds of millions of tag instances, encompassing a broad spectrum of styles and themes. This extensive coverage significantly enhances the ability of models to learn and generate complex styles. Additionally, datasets such as Comic Faces and Cartoon Faces provide paired synthetic comic facial images, further enriching the research landscape of anime style transfer by offering diverse and stylized facial representations.

In the realm of video style transfer, datasets like UADFV\cite{yang2018exposingdeepfakesusing}, EBV\cite{li2018ictuoculiexposingai}, and Deepfake-TIMIT\cite{korshunov2018deepfakesnewthreatface} provide a substantial number of video samples that cover a variety of style transfer tasks, including facial expression swapping and deepfake detection. These datasets not only boast large volumes of data but also exhibit significant diversity and authenticity, presenting models with a wide range of training samples and challenging testing environments. The diversity and real-world applicability of these video datasets are crucial for developing robust and generalizable video style transfer models.

Text style transfer datasets are equally comprehensive and varied. Datasets such as the YAFC Corpus\cite{rao2018dearsirmadami}, StylePTB\cite{lyu2021styleptbcompositionalbenchmarkfinegrained}, and TextBox\cite{tang-etal-2022-textbox} encompass a wide array of text style tasks, ranging from sentiment transfer to politeness adjustments. These datasets typically include large pairs of texts or pseudo-parallel data generated through automated methods, providing ample training resources for models to generate text with varying styles while preserving the original content. Moreover, specialized datasets like ParaDetox\cite{logacheva-etal-2022-paradetox}, which focus on detoxification tasks, offer significant support for enhancing the safety and politeness of models in practical applications, ensuring that generated text adheres to desired ethical standards.

In the field of 3D style transfer, datasets such as 100STYLE\cite{mason2022local} and Bandai-Namco-Research\cite{kobayashi2023motion} aggregate extensive collections of 3D models and animation frames, addressing diverse style conversion needs. These datasets not only provide large quantities of data but also maintain high quality and diversity, offering varied samples and challenging tasks that are essential for training and evaluating 3D style transfer models. The richness and variability of these 3D datasets enable models to effectively learn and reproduce complex stylistic transformations in three-dimensional spaces.

\subsection{Data Augmentation}
\label{Data Augmentation}
Data augmentation has emerged as a crucial technique in enhancing the performance and robustness of style transfer systems. The evolution of data augmentation methods in this domain has shown significant progress since 2019, with researchers exploring various innovative approaches to expand and diversify training datasets.

Early contributions in this field include STaDA \cite{Zheng_2019}, which demonstrated the effectiveness of utilizing style transfer techniques themselves as a means of data augmentation. In the same year, Bao et al. \cite{Bao2019CycleGANBasedES} introduced a CycleGAN-based emotion augmentation approach that specifically targeted improvements in speech emotion recognition, showcasing the versatility of style transfer augmentation across different modalities. The cross-domain neural style transfer augmentation method presented by Xu et al. \cite{xu2019crossdomainimageclassificationneuralstyle} further expanded the possibilities by enabling effective transfer learning between different domains.

The field witnessed several fundamental augmentation techniques that, while not specifically designed for style transfer, have proven valuable in this context. Notable among these are AugMix \cite{hendrycks2020augmixsimpledataprocessing}, which introduced a systematic approach to improving model robustness and uncertainty estimation, and CutMix \cite{yun2019cutmix}, which innovatively combined image patches and their corresponding labels to enhance classifier performance. InstaBoost \cite{fang2019instaboostboostinginstancesegmentation} contributed to this ecosystem by introducing an instance-aware copy-pasting method that has shown particular utility in segmentation tasks related to style transfer.

Moving into 2020, the field saw more sophisticated approaches emerge. Zhang et al. \cite{zhang-etal-2020-parallel} introduced Parallel Formality Augmentation, which specifically addressed the challenges of formality style transfer, demonstrating how targeted augmentation strategies can enhance specific aspects of style transformation. The introduction of Albumentations by Buslaev et al. \cite{Buslaev_2020} provided researchers with a comprehensive library of augmentation techniques, offering flexible tools for improving model training in style transfer applications.

Exemplified by AugLy \cite{papakipos2022auglydataaugmentationsrobustness} in 2022, have focused on creating more comprehensive and robust augmentation frameworks that work across multiple modalities. This trend reflects the growing recognition that effective style transfer systems often require diverse and multi-modal training data to achieve optimal performance.

These developments in data augmentation have collectively contributed to more robust and versatile style transfer systems. The progression from simple augmentation techniques to more sophisticated, multi-modal approaches reflects the growing understanding of the importance of diverse and high-quality training data in achieving effective style transfer results.

\end{document}